# The Mechanisms of Codon Reassignments in Mitochondrial Genetic Codes

Supratim Sengupta (1,2), Xiaoguang Yang (1), Paul G. Higgs (1)

(1) Dept. of Physics and Astronomy, McMaster University, Hamilton, Ontario L8S 4M1, Canada.
(2) Dept. of Physics and Atmospheric Science, Dalhousie University, Halifax, Nova Scotia B3H 3J5, Canada.

## Abstract

Many cases of non-standard genetic codes are known in mitochondrial genomes. We carry out analysis of phylogeny and codon usage of organisms for which the complete mitochondrial genome is available, and we determine the most likely mechanism for codon reassignment in each case. Reassignment events can be classified according to the gain-loss framework. The 'gain' represents the appearance of a new tRNA for the reassigned codon or the change of an existing tRNA such that it gains the ability to pair with the codon. The 'loss' represents the deletion of a tRNA or the change in a tRNA so that it no longer translates the codon. One possible mechanism is Codon Disappearance, where the codon disappears from the genome prior to the gain and loss events. In the alternative mechanisms the codon does not disappear. In the Unassigned Codon mechanism, the loss occurs first, whereas in the Ambiguous Intermediate mechanism, the gain occurs first. Codon usage analysis gives clear evidence of cases where the codon disappeared at the point of the reassignment and also cases where it did not disappear. Codon disappearance is the probable explanation for stop to sense reassignments and a small number of reassignments of sense codons. However, the majority of sense to sense reassignments cannot be explained by codon disappearance. In the latter cases, by analysis of the presence or absence of tRNAs in the genome and of the changes in tRNA sequences, it is sometimes possible to distinguish between the Unassigned Codon and Ambiguous Intermediate mechanisms. We emphasize that not all reassignments follow the same scenario and that it is necessary to consider the details of each case carefully.

Contact: **higgsp@mcmaster.ca**

## Introduction - Distinguishing Possible Mechanisms of Codon Reassignment

Now that many complete genomes of organisms and organelles are available, there is ample evidence that the genetic code is not as universal (Knight *et al.* 2001a; Yokobori *et al.* 2001; Santos *et al.* 2004) as previously believed (Crick 1968). Many cases are now known where a codon (or related group of codons) has been reassigned from one amino acid to another, from a stop to an amino acid, or from an amino acid to a stop. If a change in the translation system occurs in an organism such that a codon is reassigned, most of the occurrences of this codon will still be at places where the old amino acid was preferred. We would expect the changes causing the codon reassignment to be strongly disadvantageous and to be eliminated by selection. It is possible for mutations to cause disappearance of the codon in its original positions and reappearance in positions where the new amino acid is preferred. Mutations throughout the genome are required for it to readjust to the change in the genetic code. The problem is therefore to understand how codon reassignments can become fixed in a population despite being apparently deleterious in the intermediate stage before the genome has time to readjust.

Various mechanisms have been proposed to explain the process of codon reassignment. We have recently shown (Sengupta & Higgs, 2005) that these mechanisms can be described within a framework that we call the gain-loss framework. 'Gain' refers to the gain of a new tRNA gene that is able to translate the reassigned codon as a different amino acid, or the gain of function of an old tRNA gene (*e.g.* by base modification in the anticodon) so that it translates the reassigned codon in addition to the codons it previously interacted with. 'Loss' refers to the deletion of an existing tRNA for the reassigned codon, or the loss of function of the tRNA so that it can no longer translate this codon. We identified four mechanisms in the gain-loss framework, as described below

*Codon Disappearance (CD) mechanism* - This was originally proposed by Osawa and Jukes (1989, 1995). For an amino acid (or stop) with more than one codon, it is possible for all occurrences of a codon to be replaced by synonymous codons, so that the first codon disappears entirely from the genome. After this, the gain and loss in the translation system are neutral changes that do not affect the organism. After the gain and loss occur, the codon may reappear in the genome by mutations at sites where the new amino acid is preferred. The distinguishing element of this mechanism is that the codon disappears first, and the gain and loss occur during a period in which the codon is absent. For the other three mechanisms described here, the codon does not need to disappear before the change.

*Ambiguous Intermediate (AI) mechanism* - This was proposed by Schultz and Yarus (1994, 1996). They argued that a codon does not need to disappear in order to be reassigned and proposed that there is a transient period when the codon is ambiguously translated as two distinct amino acids. In terms of our gain-loss framework, this corresponds to the case where the gain occurs before the loss, *i.e.* there are two different tRNAs specific to the amibiguous codon during the intermediate period, and the new code becomes established when the old tRNA is lost.

*Unassigned Codon (UC) mechanism* - This mechanism arises as a natural possibility in our gain-loss framework (Sengupta & Higgs, 2005). It corresponds to the case where the loss occurs before the gain. There is an intermediate period where there is no tRNA available that can efficiently translate the codon; hence we say the codon is unassigned. The new code becomes established when the gain in function of the new tRNA occurs and the codon is reassigned to the new amino acid. If a codon were truly unassigned, and no tRNA could translate it at all, then the loss of the original tRNA would be lethal if the codon had not previously disappeared. However, several cases are known where an alternative tRNA is able to translate a codon (albeit less efficiently) after the tRNA that was specific to that codon has been deleted (Yokobori *et al.* 2001). Deletion of tRNAs appears to be frequent in mitochondrial genomes and we argue below that deletion of a tRNA is the prime event that instigated several of the mitochondrial codon reassignments.

*Compensatory Change (CC) mechanism* - The final mechanism that occurs in the gain-loss framework is referred to as compensatory change because of its analogy with compensatory mutations in molecular evolution. Kimura (1985) considered a pair of mutations such that each is deleterious when it occurs alone, but when both occur together they are neutral, *e.g.* in the paired regions of RNA secondary structures (Higgs 1998, 2000; Savill *et al.* 2001). The gain and the loss in codon reassignment are changes in two different parts of a genome that form a compensatory pair. It is possible that one of these changes occurs but remains infrequent in the population until the second change occurs in an individual that already has the first change. Once the gain and loss are present in the same individual, they can spread simultaneously through the population, although they did not occur at the same time. In the CC case there is no point at which individuals with ambiguous or unassigned codons are frequent in the population.

We showed using a population genetics simulation (Sengupta & Higgs, 2005) that all four mechanisms can occur within the same model depending on the parameter values. Here, we consider which mechanisms occur in real cases. We will limit ourselves to mitochondrial changes because most of the observed changes are in mitochondria and because the availability of substantial numbers of complete mitochondrial genomes makes it possible to pinpoint the changes and to study codon usage and tRNA gene content in the genomes before and after the reassignment. Studies of this type were carried out several years ago by Knight *et al.* (2001a, 2001b) and more recently by Swire *et al.* (2005). We compare our results with these previous surveys in the discussion section.

In order to interpret the changes in the genetic code, phylogenetic trees are required. We split our set of species into groups that are consistent with previous studies. Figure 1 shows fungi and related species, and Figure 2 shows plants/algae and related species. These were obtained by our own phylogenetic analysis of mitochondrial genes. Figure 3 shows alveolates/stramenopiles and related species, and Figure 4 shows metazoa. These were obtained using combined information from other sources. Details of phylogenetic methods are given in the supplementary information.

**Reassignments that can be explained by codon disappearance**

<u>Reassignments of UGA from Stop to Trp</u>

The UGA Stop to Trp change is the most frequently occurring reassignment known. The review of Knight *et al.* (2001a) lists 6 of these in mitochondria (and 4 in nuclear genomes). Our updated analysis of the mitochondrial data identifies the following 12 cases in mitochondria.
(i) Metazoa, *Monosiga* and *Amoebidium* - This change is shared by all Metazoa and by their two closest known relatives (Fig 1 and Lang *et al.* 2002).
(ii) *Acanthamoeba* - Fig 1 and Burger *et al.* (1995).
(iii) Basidiomycota (*Crinipellis* and *Schizophyllum*) - Fig 1.
(iv) Ascomycota (group containing *Penicillium* and relatives) - Fig. 1.
(v) Ascomycota (group containing *Yarrowia* and relatives) Knight *et al.* (2001a) list a single change in the ancestor of Metazoa, *Acanthamoeba* and Fungi. However, Figure 1 shows that UGA remains a stop codon in Chytridiomycota, Zygomycota and *Dictyostelium*. Therefore, it is likely that cases (i), (ii) and (iii) are separate reassignments. The reassignment has also occurred in almost all the Ascomycota, with the exception of the *Schizosaccharomyces* group. This implies that cases (iv) and (v) in the Ascomycota are also separate reassignments. These conclusions depend on the argument that a reversal of this change (from Trp to Stop) is very unlikely, which we discuss below.
(vi) Rhodophyta (*Chondrus*, *Porphyra*) - Fig. 2 and Burger *et al.* (1999).
(vii) *Pedinomonas* - Fig. 2 and Turmel *et al.* (1999).
(viii) Haptophytes - This change was reported in a subgroup of Haptophytes including *Phaecocystis* and *Isochrisis* by Hayashi-Ishimaru et.al. (1997). The complete genome of *Emiliana* (Sanchez-Puerta *et al.* 2004) shows that it also possesses the reassignment. This is consistent with our phylogeny (Fig 3).
(ix) Ciliates (*Paramecium, Tetrahymena*) - Knight *et al.* (2001) place this change at the base of the alveolates. However, our analysis of codon usage and sequences from *Plasmodium* species discussed below shows that UGA remains a Stop codon in *Plasmodium*. Thus the reassignment is not shared by all alveolates.
(x) *Cafeteria* - Fig 3.
(xi) Bacillariophyta (*Skeletonema*, *Thalassiosira*) - Fig 3 and Ehara *et al.* (2000)
(xii) Kinetoplastida (*Trypanosoma, Leishmania*) - Inagaki *et al.* (1998). (Not shown on figures).

In species in which only the UGG codon codes for Trp, the tRNA-Trp has a CCU anticodon. Mutation of the wobble position C to a U creates a UCU anticodon that can pair with both UGA and UGG: hence the reassignment of UGA to Trp. We have analyzed the tRNA-Trp sequences of all available mitochondrial genomes to determine in which species this mutation has occurred. We find that all species in which UGA is Stop have a CCU anticodon, as expected. Almost all species in which UGA is Trp have a UCU anticodon, but we find some exceptions below. Table 1 shows the number of occurrences of each of the standard Stop codons together with the tRNA-Trp anticodon. When UGA is Stop it is usually used less frequently than the preferred stop codon UAA. When the reassignment occurs, many UGG Trp codons mutate synonymously to UGA; hence UGA becomes frequent. This is seen in all species having a UCA anticodon. However, UGA is also frequent in *Amoebidium*, *Crinipellis* and *Schizophyllum*, which have a CCA anticodon. Therefore there must be a post-translational modification of the C base in the anticodon that permits translation of UGA, rather than a mutation in the gene. Another special case is the Kinetoplastids. These have no mitochondrial tRNAs, and import all the required tRNAs from the nucleus. The tRNA-Trp from the nucleus has CCA anticodon because the canonical code is used in the nucleus. The same tRNA is imported to the mitochondrion and then undergoes a base modification so that it can translate both UGA and UGG (Alfonzo *et al.*, 1999).

The predictions from our theory and simulations (Sengupta and Higgs, 2005) were that Stop codons are most likely to be reassigned via the CD mechanism because they are rare in the first place (and chance disappearance is therefore relatively likely), and also because, if the codons do not disappear, the penalty for read-through of a stop codon is likely to be larger than the penalty for mis-translation of an amino acid. (However, read-through of a stop codon is not necessarily lethal, especially if the number of additional codons translated until the next random occurrence of a stop codon is not too large.) The figures in Table 1 make a strong case that the CD mechanism is responsible. UGA is rare in almost all species where it is used as a Stop codon. Many of these genomes have high AU content, which is probably the reason UAA is preferred over UGA as a Stop codon. Many of the closest relatives to the species where UGA is reassigned have particularly low usage of UGA, *e.g. Allomyces* and *Rhizopus*, close relatives of the Basidiomycota, have 0; *Cyanidioschyzon*, close relative of *Chondrus* and *Porphyra*, has only 2; *Scenedesmus* and *Chlamydomonas*, close relatives of *Pedinomonas*, have 1 and 0 respectively; and *Plasmodium*, close relative of *Paramecium* and *Tetrahymena*, has 0. This last example shows that UGA is not reassigned in *Plasmodium*, as mentioned in case (ix) above. These examples clearly show that disappearance of UGA is possible, and that in many of the cases, there is good evidence that UGA was absent or almost absent, at the time it was reassigned.

A mutation pressure from GC to AU, which is implicated in the disappearance of UGA, will also tend to cause rapid mutations from UGG Trp codons to UGA after the Trp tRNA gains the ability to decode UGA. This is one reason that reversal of the change is unlikely. There are often around 100 UGAs in genomes where the reassignment has occurred, and it would be very difficult for this large number to disappear by chance because this would act against the mutation pressure. A second reason is that the reassignment to Trp would be associated with the loss of function of the release factor that originally interacted with the UGA. A reversal would also require regaining of the function of the release factor.

The unlikeliness of the reversal of the reassignment is important in our interpretation of cases (i) to (v) above. The codon usage and tRNA-Trp anticodon data (Table 1) show that UGA is not established as a Trp codon in *Dictyostelium*, Chytridiomycota, Zygomycota, or *Schizosaccharomyces.* This means that multiple reassignment events are required to explain the observed pattern. It appears that UGA was rare and prone to disappearance right from the base of the Metazoa/Fungi tree in Figure 1, and the release factor may already have lost its function. Therefore all that is required for the reassignment to occur is the simple mutation of the tRNA anticodon. In groups where the release factor was lost, but the tRNA mutation did not occur, the UGA codon would be effectively unassigned and would be selected against. Seif *et al.* (2005) argue that this has occurred in *Mortierella* and *Schizosaccharomyces*, where there is evidence of a small number of UGA acting as Trp codons that are translated very inefficiently by the standard Trp tRNA with CCA anticodon. A genetic code change might easily become established in these species in the future if the mutation occurred in the tRNA.

## The probability of disappearance of UGA codons

It is possible to calculate the probability $P_{dis}$ that a codon will disappear using a simple model of the mutation process. Swire *et al.* (2005) have used this method to show that in most cases of stop codon reassignment, the probability of disappearance of the codon was relatively large, whereas is many cases on sense codon reassignments, the probability was extremely small. Hence they argue that stop codons were reassigned via CD but sense codons were not. We agree with this conclusion in almost all cases, but there are a few cases where we will argue for CD in sense codon reassignments as well.

In this section, we consider only UGA stop codons. The probability of UGA codon disappearance can be calculated as follows. Let the equilibrium frequencies of the bases under the mutational process be $\pi_A$, $\pi_C$ , $\pi_G$ and $\pi_U$. Let $f_{UGA}$, $f_{UAA}$, and $f_{UAG}$ be the relative frequencies of the three stop codons ($f_{UGA} + f_{UAA}, + f_{UAG} = 1$). If there is no selective preference of one stop codon over another, we expect these frequencies to be in equilibrium under mutation. Therefore

$$\frac{f_{UGA}}{f_{UAA}} = \frac{f_{UAG}}{f_{UAA}} = \frac{\pi_G}{\pi_A}.$$

Hence $f_{UGA} = f_{UAG} = \pi_G / (2\pi_G + \pi_A)$. In mitochondrial genomes, the two strands of the genome are not equivalent, the four base frequencies are all different, and it is not true that $\pi_C = \pi_G$ and $\pi_A = \pi_U$ (see Urbina *et al.* 2006). The values of the frequencies can be estimated from the frequencies of the bases at fourfold degenerate (FFD) sites. These are calculated by summing over all third-position sites that are FFD. If the total number of stop codons in the genome is $N_{stop}$ then the probability of disappearance of UGA is $P_{dis} = (1 - f_{UGA})^{N_{stop}}$.

We do not know the values of $N_{stop}$ or $f_{UGA}$ at the point where the codon reassignment occurred, but we can use the species that are close to the reassignment point on the tree as an estimate, as shown in Table 2. $N_{stop}$ is the the sum of UAA and UAG from Table 1. UGA codons do not contribute to $N_{stop}$ because these are now Trp codons in these species. We used *Monosiga* as a proxy for case (i). The resulting probability is high ($P_{dis}$ = 0.17), and is consistent with the CD mechanism. *Amoebidium* was not used in this case because it has a derived multi-chromosome structure of the mitochondrial genome, and metazoa were not used because they have a much reduced genome size. Both of these would be poor estimators of what the genome was like at the reassignment point. In all these species, $\pi_G$ is much less than $\pi_A$, but the bias is more extreme in some species than others. In case (ix), if either of the *Tetrahymena* species is used as proxy, a fairly high $P_{dis}$ is obtained, but if *Paramecium* is used, $P_{dis}$ is much lower. Similarly in case (iii), $P_{dis}$ is high if we use *Schizophyllum*, but low if we use *Crinipellis*. These two species differ in base frequencies and also in the number of genes on the genome (*i.e.* $N_{stop}$). Although $N_{stop}$ = 89 in *Crinipellis* (also known as *Moniliophthora perniciosa*), this genome has many genes labelled as 'hypothetical protein' that do not have homologues in related species. These genes may be recent insertions, or may not even be expressed sequences. Thus, $N_{stop}$ was probably much less than 89 at the time of the reassignment, and $P_{dis}$ would be correspondingly higher. In both these cases, it is reasonable to conclude that the properties of the genome at the point of reassignment were such that codon disappearance was not too unlikely.

The only other case where $P_{dis}$ is low is case (ii), where $P_{dis} = 6.4 \times 10^{-5}$ if *Acathamoeba* is used as proxy. There is no other related species for which data is available. We have already seen that fluctuations in base frequencies can change $P_{dis}$ by several orders of magnitude. Therefore a value of order $10^{-5}$ does not appear ridiculously small, and the true value could well have been much higher at the point the reassignment actually occurred. Finally we comment in case 17 of Swire *et al.* Table 2a, where they estimate a low probability of $P_{dis} = 2.7 \times 10^{-11}$. This corresponds to our case (vi), where we estimate $P_{dis} = 3.6 \times 10^{-2}$ if *Chondrus* is used as proxy and $P_{dis} = 1.4 \times 10^{-3}$ if *Porphyra* is used. Both of these are much higher than the estimate of Swire *et al.* We also note that, according to our phylogeny, these two species are related and we assume a single reassignment in their ancestor, whereas Swire *et al.* treat them as independent (cases 16 and 17). The change in the phylogeny would not affect our estimate of $P_{dis}$. Thus, we conclude that all these examples of UGA codon reassignment are consistent with the CD mechanism. (Cases (xi) and (xii) are not included in Table 2 because there is insufficient data to perform the calculation.)

Reassignments of UAG stop codons

Two cases of reassignment of UAG from Stop to Leu are known: one in the two chytrids, *Rhyzophidium* and *Spizellomyces* (Fig. 1, Laforest *et al.* 1997), and the other in *Scenedesmus* (Fig. 2; Hayashi-Ishimaru *et al.* 1996; Kück *et al.* 2000). In Table 1, UAG is also seen to be rare in general, and rare particularly in the close relatives of the species reassigned: only 2 in the other chytrid, *Hyaloraphidium*; and 0 in *Pedinomonas*, relative of *Scenedesmus*. These reassignments show the same pattern as the UGA Stop to Trp examples, and can also be attributed to the CD mechanism. Disappearance of UAG would also be favoured by mutation pressure increasing the AU content.

The reason UAG is reassigned less frequently than UGA may be because of the relative difficulty of the required change in the tRNA. In the case of UGA, the existing tRNA-Trp can simply mutate its anticodon. In the UAG case it is necessary to create a new tRNA-Leu with a CUA anticodon to pair with UAG. This can be done by making a mutation in the second position of a tRNA-Leu with CAA anticodon. Such a mutation in the anticodon would not change the amino acid specificity of the tRNA because the anticodon of the tRNA-Leu does not act as an identity element for recognition by the leucyl-tRNA synthetase (Asahara et.al. 1993), in contrast to most other amino acyl-tRNA synthetases, which recognize the anticodon. However, this can only be done after duplication of the tRNA-Leu, because one copy of this gene is still required to translate UUG Leu codons. Thus, this reassignment is relatively rare because it requires a prior tRNA duplication. It is interesting to note that in most genomes, both UUA and UUG would be Leu, and there would be a single tRNA with UAA anticodon to translate both these. Another peculiarity of *Scenedesmus* is that UUA codons are not used (Kück *et al.* 2000), but UUG codons are frequent. This is clearly related to the fact that the tRNA-Leu has CAA anticodon in this species.

UAG has been reassigned to Ala in a group of green algae that are closely related to *Scenedesmus* (Hayashi-Ishimaru *et al.* 1996). Complete genomes are not available for this group, therefore they are not included in our Figure 2. It seems likely that UAG disappeared in the ancestor of *Scenedesmus* and the other algae, and that the codon was captured by Leu in *Scenedesmus* and by Ala in the other species.

Table 2 also shows the probability of disappearance of the UAG codon for these two cases of UAG reassignment. This is calculated in the same way as for UGA above. These values do not appear unreasonably small in view of the uncertainty in the genome properties at the point of reassignment. Therefore, we conclude that these changes are consistent with the CD mechanism.

Sense codon reassignments linked to codon disappearance

It is not only Stop codons that can be reassigned by the CD mechanism. The following examples show sense codon changes occurring via this mechanism. In the canonical code, both CUN and UUR code for Leu and these two codon blocks are accessible to one another by a synonymous mutation at the first position. Similarly CGN and AGR both code for Arg and are also accessible to one another via a synonymous mutation at first position. In AU rich genomes, CUN and CGN codon families are subject to disappearance and replacement by UUR and AGR. Table 3 gives the number of occurrences of the codon groups for Leu and Arg in several Fungi. Base frequencies at FFD sites are also shown. C and G frequencies are very low in all the species in Table 3, indicating a strong mutation pressure towards A and U.

The species in the top half of the table (*P. canadensis* and above) use CUN for Leu, as in the canonical code. These species possess a tRNA-Leu with anticodon UAG for this family. In all these cases the CUN family is less frequent than the UUR family. The six species labelled (a) in Table 3 have undergone a reassignment of the CUN family to Thr. These species possess an unusual tRNA-Thr with anticodon UAG (Sibler *et al.* 1981, Osawa *et al.* 1990) and the usual tRNA-Leu(UAG) is not found in the genome. In *K. lactis*, the codon family is not used at all. Significantly, this is the only species in which no tRNA exists that pairs with this codon family.

The fact that the tRNA-Thr(UAG) appears only in species where the tRNA-Leu(UAG) is absent suggests that the new gene evolved directly from the old tRNA-Leu(UAG) gene. This requires no change in the anticodon, but changes elsewhere are required in order that the tRNA is recognized by the threonyl-tRNA synthetase and not the leucyl-tRNA synthase. Although many amino acyl-tRNA synthetases recognize the anticodon, recognition of tRNA-Leu by leucyl tRNA-synthetase is exclusively determined by the bases in the large variable arm (Asahara et.al. 1993). Changes in the variable arm could have prevented the tRNA acting as a tRNA-Leu and would have left it open to evolving a new function as a tRNA-Thr.

The codon usage figures in Table 3 indicate that the change is attributable to the CD mechanism. The number of occurrences of CUN was most likely driven to zero prior to the branching of *A. gossypii* (see Fig. 1). A major change in the tRNA-Leu(UAG) gene was then possible allowing it to be charged with Thr. CUN codons later reappeared with a new meaning. In the context of our gain-loss framework (Sengupta and Higgs, 2005), this example is interesting in that the gain and loss of function occur in the same tRNA.

Although it is likely that the unusual tRNA-Thr in these species evolved directly from the old tRNA-Leu, the alternative scenario is that there was a duplication of the tRNA-Thr(UGU) gene that decodes the normal ACN Thr codons. One of these genes might then have changed its anticodon to UAG (two mutations required). In order to determine the origin of the tRNA-Thr(UAG) gene, we constructed a phylogeny of all tRNAs from *P. canadensis*, *K. thermotolerans* and *S. castellii*, *i.e.* from representative species before and after the codon reassignment. The tRNAs of each type formed monophyletic triplets, with the exception of the genes with UAG anticodons. The tRNA-Thr(UAG) genes in *K. thermotolerans* and *S. castellii* showed no close relationship to either the tRNA-Thr(UGU) genes or the tRNA-Leu(UAG) gene in *P. canadensis*. Thus, neither of the two possible scenarios was supported. Despite this, it stills seems most likely that the new tRNA-Thr gene evolved from the old tRNA-Leu, a conclusion also reached by Sibler *et al.* (1981) and Osawa *et al.* (1990).

One possibility that cannot entirely be ruled out from the codon usage data is that the changes in the tRNA-Leu(UAG) occurred when the CUN codons were very rare but not entirely absent. The changes might have been such as to immediately cause the tRNA to be charged by Thr, or might have happened more gradually, so that the same tRNA could be ambiguously charged by Leu and Thr during the changeover period. We would then have to count this as an example of the AI mechanism. Nevertheless, the low CUN number resulting from mutation pressure is clearly a major factor in this case, and we therefore feel comfortable in classifying it under the CD mechanism. The tRNA gene has undergone considerable modification, including an unusual insertion that makes the anticodon loop larger than the standard 7 bases. In our opinion, it is unlikely that such a large change could have happened whilst the tRNA remained simultaneously functional for both amino acids.

Table 3 also shows disappearances in the Arg codons that happen in parallel with those in the Leu codons. Mutation pressure away from C causes replacement of CGN codons by AGR. Once CGN codons have disappeared, there is no penalty to the deletion of the tRNA-Arg(ACG) gene. This gene is absent in the species indicated (c) in Table 3, and CGN is absent in these species (or apparently just 1 in *C. glabrata*). On the other hand the gene is still present in *S. cerevisiae* and *K. thermotolerans*, and a handful of CGN codons still remain in these species. The tRNA-Arg(ACG) genes in *S. cerevisiae* and *K. thermotolerans* show clear sequence homology to that in *P. canadensis*. From the phylogeny of these species in Figure 1, we see that it requires 4 independent deletions of the tRNA-Arg(ACG) to explain the observed pattern of absences of the gene, indicating that deletion of redundant genes is a rapid phenomenon.

Figure 1 shows several separate reassignments of CGN from Arg to unassigned and only one reassignment of CUN from Leu to Thr. This is because the symbols label points where the tRNAs were deleted or mutated, not points where the codon disappeared. Our interpretation is that both CUN and CGN codons were absent in the ancestor of the group of yeast species prior to the branching of *A. gossypii*, and that change in the tRNA-Leu occurred at this point, whereas the deletions of the tRNA-Arg genes did not occur till after the split between the species. This would permit small numbers of CGN codons to reappear in some lineages where the gene was not deleted (*e.g. S. cerevisiae* and *K. thermotolerans*). The final twist in this story is that the CUN codons are absent in *K. lactis*. This means that there must have been a further reassignment of CUN from Thr to unassigned in this species only. Clearly the codons have disappeared in this case, which could be due to drift, or because of poor functioning of the tRNA-Thr(UAG) gene in this species, which would have caused selection against this codon block. We have little evidence in this case, but for the purposes of our summary table in the discussion section below (Table 6), we classed it as CD.

Table 3 shows another example of the disappearance of the CGN block in *Y. lipolytica*. The corresponding tRNA-Arg is also deleted in this species (Kerscher *et al.* 2001). In *S. pombe* and *C. stellata*, the species that branch before and after *Y. lipolytica*, the tRNA-Arg is still present and the CGN codons are still used. CGN has thus become unassigned in this single species. This also appears to be a result of mutation pressure causing disappearance of CGN before deletion of the tRNA-Arg.

The probability of disappearance of CUN and CGN codons

The numbers of Leu and Arg codons in these genomes is much larger than the number of stops. Therefore, it is more difficult for these sense codons to disappear than it is for stops. Nevertheless the probability of codon disappearance is not unreasonably small in these cases. If we assume that all six Leu codons are in equilibrium, we find

$$\frac{f_{CUX}}{f_{CUG}} = \frac{\pi_X}{\pi_G} \text{ (for X = A, C or U)},$$

$$\frac{f_{UUG}}{f_{CUG}} = \frac{f_{UUA}}{f_{CUA}} = \frac{\pi_U}{\pi_C},$$

from which the sum of the relative frequencies of all the CUN block codons is

$$f_{CUN} = \pi_C / (\pi_C + \pi_U(\pi_A + \pi_G)).$$

The probability of disappearance of the CUN block is $P_{dis} = (1 - f_{CUN})^{N_{Leu}}$, where $N_{Leu}$ is the number of Leu codons in the genome.

However, Table 3 shows that this calculation is hardly necessary. In *A. gossypii*, the species branching immediately after the proposed point of the codon reassignment, %C = 0 at FFD sites, *i.e.* the mutational bias against C is so strong that there are no C bases at all in any FFD site. Thus $f_{CUN}$ = 0, and $P_{dis}$ = 1. Presumably the chance of mutating to a C is not precisely zero in *A. gossypii*, but it is small enough that all the FFD C-ending codons disappear. Therefore it is not unreasonable that the CUN block should have disappeared in the ancestor of this group.

As we do not know that %C was exactly zero in the ancestor, we would like to know how small it must have been in order for there to be a significant probability of disappearance of CUN. It can be seen that %C is low in all species in Table 3, but variations of plus or minus 1% make a big difference in estimation of $P_{dis}$. Estimation of base frequencies is further complicated by the presence of intronic ORFs in many of these species. It has previously been shown that codon usage patterns differ between the intronic ORFs and the other genes (Bullerwell *et al.* 2003; Talla *et al.* 2005). For several species in Table 3, we have listed separate codon information for the genes excluding the intronic ORFs (E) and for the full set of genes including the intronic ORFs (I). In every case considered, %C is slightly higher if the intronic ORFs are included. These ORFs vary in number between species, and are not present in some species, including *A. gossypii*. This makes it difficult to know the number of genes in the ancestral genome at the point of CUN reassignment. It also suggests that these sequences may be recent insertions and that the base composition may not have adjusted to the equilibrium base composition of the rest of the genome. Furthermore, the presence of rare codons in the intronic ORFs suggests that these sequences my not be expressed at a high level, or that their expression may not be important for the organism. For example, in *K. lactis* and *K. thermotolerans*, AUA is avoided because of deletion of the corresponding tRNA (to be discussed below), but there are substantially more AUA codons in the intronic ORFs than the regular genes. For these reasons, it is likely that the base frequencies in the genes excluding the intronic ORFs are better indicators of the equilibrium base frequencies of the mutational process affecting the most important genes in the genome. We also note that the presence of the intronic ORFs is sometimes associated with alternative splicing, so that the same exon is used in several proteins (e.g. *S. cerevisiae* and *C. stellata*). Clearly we do not want to count the same exon twice when counting codons. A full investigation of all these effects on intronic ORFs is beyond the scope of this paper.

Given all the caveats above, it is clear that any estimate of $P_{dis}$ is subject to a very large uncertainty. However, we wish to give a fair estimate for a case where %C is not exactly zero. We will use the figures for *C. glabrata* because the base frequencies with and without the intronic ORFs differ less than in most species. To be conservative, we will use the figures including the ORFs. From Table 3, we obtain $f_{CUN} = 0.0354$, and $P_{dis} = (0.9646)^{415} = 3.2 \times 10^{-7}$. Slight reductions in %C would increase this figure by orders of magnitude. We conclude the bias against C must have been at least as strong as it is in *C. glabrata* (i.e. %C < 0.9) in order for the CUN codons to have a reasonable chance of disappearing. The fact that %C becomes exactly zero in *A. gossypii* and is less than 0.9% in several other cases in Table 3 shows that these extreme biases do occur. Swire *et al.* (2005) have calculated many examples of codon disappearance probabilities. The CUN reassignment corresponds to case 3 of Swire *et al.* (2005) (Table 2b). For some reason, they estimated a vanishingly small probability of disappearance of the CUN codons ($P_{dis} = 6.8 \times 10^{-427}$), which is the smallest of any of the cases they list. They thus argued that codon disappearance was impossible in this case. We disagree, since we have shown that $P_{dis}$ is high due to the extreme bias against C bases. This case is exceptional, however. In all the other examples of sense to sense codon reassignments discussed by Swire *et al.,* we agree that the probability of codon disappearance is indeed unrealistically small and that the reassignment must have occurred without codon disappearance.

Using an argument similar to that above, the relative frequency of CGN codons is $f_{CGN} = \pi_C / (\pi_C + \pi_A(\pi_A + \pi_G))$, and the probability of disappearance of the CGN block is $P_{dis} = (1 - f_{CGN})^{N_{Arg}}$. We have already seen that %C = 0 in *A. gossypii* and is very low in related species. $N_{Arg}$ is much lower than $N_{Leu}$, so disappearance of CGN in the ancestor of this group is therefore quite likely, and is easier than disappearance of CUN. For the case of *Y. lipolytica*, if we use the figures from this species as a proxy for the ancestor, we obtain $P_{dis} = (1\text{-}0.0426)^{75}$ = 0.038, which is not unreasonably small.

Clearly it is much more difficult for sense codons to disappear than stop codons, as the number of codons involved is much larger. For this reason, we need an extremely strong base frequency bias in order to cause disappearance. However, in the above cases we have shown that the bias is indeed extreme, and that the CD mechanism seems likely. A reviewer of this paper has pointed out that Osawa and Jukes (1989), the original proponents of the codon disappearance model, insisted that, according to their theory, every single occurrence of the codon must disappear from the genome before the reassignment occurred. From current genome data it will never be possible to tell whether every occurrence of the codon disappeared at the time the reassignment occurred. Therefore, if we insist on this strict definition, it is impossible to assign a mechanism to any of the reassignment events. However, to be more constructive and more practical, we wish to make a clear distinction between the examples discussed above and those that we will discuss in the remainder of the paper. In the above examples, there is a clear case that codon disappearance is likely due to the direction of the mutation pressure and where the best estimates of the codon disappearance probabilities are not too small. The extreme rarity (and probably absence) of these codons was clearly a major factor in allowing the codon reassignment to occur. In contrast, in the examples below, there is clear evidence that the codon was not rare at the time of the reassignment, and that the codon reassignment was initiated by some other event not related to codon disappearance.

**Reassignments that cannot be explained by codon disappearance**

Reassignment of AUA from Ile to Met is initiated by loss of a tRNA-Ile gene.

In the canonical code, the AUN codons are unusual in that AUU, AUC and AUA code for Ile, and only AUG codes for Met. The tRNAs need to distinguish between codons ending in A and G, which is not usually necessary (except for the UGA Trp codon). The translation system in bacteria uses two tRNAs for Ile: one with anticodon GAU, which translates AUY codons and one with anticodon CAU, which translates AUA only. It is known from studies on *E. coli* (Muramatsu *et al.* 1988) that the C at the wobble position undergoes a post-transcriptional base modification to $K^2C$ (lysidine), which allows this tRNA-Ile to pair with the AUA codon. The same gene also is present in many mitochondrial genomes in which AUA is translated as Ile, and it is presumed that the same base modification occurs. However, the tRNA-Ile(CAU) gene is absent in mitochondrial genomes in which AUA is reassigned to Met. Before codon reassignment, tRNA-Met also has a CAU anticodon, which pairs with AUG only, but modification of the wobble-position C allows it to pair with both AUA and AUG (Tomita *et al.* 1999b).

In order to locate and understand the cases of AUA codon reassignment in mitochondria, it is crucial to establish which genomes contain the tRNA-Ile(CAU) gene. This is complicated by the fact that some genomes contain three tRNAs with CAU anticodons (a Met initiator, a Met elongator and an Ile), and the Ile tRNA is often mistakenly annotated as Met. To uncover mis-annotations, we constructed a phylogeny of all the tRNAs with CAU anticodon from all the fungi genomes in our data set. The genes fell into three groups that could be reliably identified. Table S3 in the supplementary information lists the position of the three tRNAs in each genome, and hence shows in which genomes there have been gene deletions.

From this, we note that the tRNA-Ile is missing from the 7 species from *A. gossypii* downward on Table 3. Hence, this gene must have been deleted prior to the branching of *A. gossypii*, as shown in Fig. 1. Before the deletion of the tRNA-Ile(CAU) gene, AUA and AUU are frequent codons for Ile but AUC is rare (*e.g. C. albicans, C. parapsiplopsis* and *P. canadensis*). This is to be expected from the FFD base frequencies. Mutation pressure is therefore in the wrong direction to cause disappearance of AUA, and in fact AUA is the most frequent Ile codon in *P. canadensis*, the closest relative to the species concerned. We therefore argue that this reassignment was initiated by loss of the tRNA-Ile while the AUA codon was frequent in the genome. Hence, it is an example of the UC mechanism, according to our classification.

Immediately after the loss of this tRNA, the AUA codon had no specific tRNA to pair with it, and it may be said to have been unassigned. The tRNA loss could not have been lethal, however, so there must have been another way of decoding AUAs. In Metazoa, some genomes are known where the tRNA-Ile(CAU) gene has been deleted, but the AUA codon has not been reassigned. Yokobori *et al.* (2001) have shown that in such cases, AUA is translated as Ile by the tRNA-Ile(GAU), even though this requires GA mispairing at the wobble position. It is presumed that this is tolerated if there is no better alternative available, *i.e.* if the tRNA-Ile(CAU) has been deleted but no modification to the tRNA-Met has occurred. This may be what is happening in the two *Kluyveromyces* species, where usage of AUA is very low, despite the fact that A is also frequent at FFD sites in these species. This demonstrates a definite avoidance of AUA, probably due to the fact that there is no tRNA that translates the AUA codon efficiently. Talla *et al.* (2005) argue that AUA is assigned to Met in *K. thermotolerans* based on only two occurrences of this codon in the *VAR1* gene. It may be more reasonable to say that AUA is unassigned in both *K. lactis* and *K. thermotolerans*. In *C. glabrata*, Koszul *et al.* (2003) conclude that AUA is associated with Met based on comparison of *C. glabrata* sequences with *S. cerevisiae* sequences from which they find that 8 out of 16 AUA codons in *C.glabrata* appear at conserved Met sites encoded by AUA in *S. cerevisiae*. There is still some avoidance of AUA codons in *C. glabrata* nevertheless. In *A. gossypii* and the *Saccharomyces* species, AUA is frequent, and we presume that it is efficiently translated as Met. This may imply that there have been two independent adaptations of the tRNA-Met to deal with AUA in these two lineages.

We wish to underline the difference between the AUA reassignment, and the reassignments of CUN and CGN in the same species (see previous section). For CUN and CGN, mutation pressure drives the disappearance. In both these cases there is only one tRNA for the block of four codons, and if this were deleted (in the CGN case) or drastically mutated (as probably happened in the CUN case), there would be no plausible alternative tRNA that could step in. Hence these changes can only occur after the codon disappears. In contrast, mutation pressure leads to frequent usage of AUA for Ile. Loss of the tRNA-Ile(CAU) is possible without codon disappearance because the tRNA-Ile(GAU), and possibly also the tRNA-Met(CAU), have some ability to pair with this codon. A calculation of the AUA codon disappearance probability also serves to highlight the difference with the previous section. If the three Ile codons are in equilibrium, $f_{AUA} = \pi_A / (\pi_U + \pi_C + \pi_A)$, and $P_{dis} = (1 - f_{AUA})^{N_{Ile}}$. Using *C. glabrata (I)* as an example, as we did for the CUN disappearance, we obtain $P_{dis} = (1-0.502)^{343} = 1.4 \times 10^{-104}$, which is indeed vanishingly small. As described in the previous paragraph, the adaptation to using AUA for Met seems to have been gradual. A possible reason for this could be that the change in the tRNA-Met is due to a base modification of the wobble-position C, not a mutation (as in the tRNA-Trp, where the wobble position C mutates to U when the UGA codon is reassigned to Trp). The difference between the avoidance of AUA seen in *Kluyveromyces* and the fairly frequent use of AUA as Met in *Saccharomyces* may lie in the fraction of tRNAs in which this base gets modified.

The other species of fungus in which tRNA-Ile(CAU) is missing is *Schizosaccharomyces octosporus*, although the gene is still present in *S. pombe* and *S. japonicus*. Table 3 shows that AUA is absent altogether in *S. octosporus* (excluding intronic ORFs), whereas comparison with the other two species suggests that there were thirty or forty AUA codons present before the loss of the tRNA. Once again, it is unlikely that these codons would have disappeared by chance, and A is a frequent base at FFD sites in all three *Schizosaccharomyces* species. Hence the disappearance of AUA in *S. octosporus* is due to selection against this codon after the loss of the tRNA. Bullerwell *et al.* (2003) show that AUA has not disappeared entirely in the intronic ORFs of *S. octosporus*. This suggests that there is weaker selection on codon usage in the intronic ORFs because these genes are less strongly expressed than the standard genes. It also confirms that selection must have driven the disappearance of AUA in the standard genes, because if mutation pressure were responsible, it would have affected the intronic ORFs equally. Thus, we conclude that the initial event must have been the deletion of the tRNA-Ile(CAU), and hence this corresponds to the UC mechanism. In this case, the codon has not been reassigned to Met, but there is the potential for this to occur in the future if there is a gain of function in the tRNA-Met.

From Table S3 in the supplementary information, we see that *C. albicans, C. parapsilosis* and *Y. lipolytica* have only one tRNA-Met. Possibly this plays the role of both initiator and elongator. We do not consider it further because this deletion is not associated with any codon reassignment.

AUA is also reassigned to Met in some Metazoa. The tRNA-Ile(CAU) gene is present in close relatives of the Metazoa, *Amoebidium* and *Monosiga* (Burger *et al.* 2003), and also in Porifera (Lavrov *et al.* 2005). AUA is Ile in these species. The gene is deleted in all remaining Metazoa (see Fig. 4), which suggests that it was lost after the divergence of the phylum Porifera from other metazoan phyla. Cnidaria are unusual in that they have lost almost all tRNAs from the mitochondrial genome, and require import of tRNAs from the nucleus. AUA remains Ile in this group, presumably because the tRNA-Ile(CAU) is imported. There are also two other groups in which AUA remains as Ile: Platyhelminthes, or more specifically Rhabditophora (Telford *et al.* 2000), and Echinoderms/Hemichordates. These phyla retain sufficient tRNAs to translate the full code and there is no suggestion of tRNA import. It therefore appears that AUA is translated as Ile by the tRNA-Ile(GAU). This is consistent with the argument of Yokobori *et al.* (2001) that GA pairing at the anticodon wobble position is tolerated if there is no better alternative anticodon-codon pairing solution available.

In the remaining metazoan phyla, AUA is reassigned to Met. The usual tRNA-Met anticodon is CAU. This can be modified in several different ways to pair with AUA in addition to AUG. In *Drosophila*, modification from C to 5-formylcytidine ($f^5C$) at the wobble position occurs (Tomita *et al.* 1999b). These authors also show that the unmodified CAU anticodon can translate both codons if the base in the 37th position of the tRNA is $t^6A$ ($N^6$-threonylcarbamoyladenosine) instead of A (see also Kuchino *et al.* 1987). Note that the 37th position is the nucleotide immediately 3' of the anticodon. In Urochordates, the tRNA-Met has a U*AU anticodon (where U* is an unknown derivative of U). This is a third means by which the tRNA-Met has gained the ability to translate AUA. There is insufficient evidence to say exactly where these modifications to the tRNA-Met evolved, and there are many species for which the details of the tRNA-Met are not known. Nevertheless, the existence of these different modifications to the tRNA-Met suggests that the reassignment of AUA to Met has occurred independently at least three times in different phyla *after* the loss of the original tRNA-Ile(CAU) gene (as shown in Figure 4). Note that the AUA codon is common in Porifera and Cnidaria, and we did not find any example of disappearance of this codon in any of the Metazoa (See Table 4). Hence, there is no suggestion that the codon disappeared before the loss of the tRNA-Ile(CAU). As the tRNA loss occurred first, this is an example of the UC mechanism, very similar to the case of AUA reassignment in Fungi.

We interpret the decoding of AUA as Ile in Platyhelminths and Echinoderms/Hemichordates as indicating that an appropriate change in the tRNA-Met did not occur, and that AUA has always been Ile in these lineages. An alternative is that AUA was assigned to Met once at the point of the loss of the tRNA-Ile(CAU) and that there were reversals of the change in these two groups (as shown in Figure 2 of Knight *et al.* (2001a) and Figure 1 of Yokobori *et al.* (2001)). However, this scenario does not explain why more than one type of modification of the tRNA-Met should exist. Also, once the AUA codon is captured by Met and mutations have occurred throughout the genome (so that AUA appears in positions where a Met rather than an Ile is required), then the organism is better adapted than it would be if AUA were inefficiently translated by the tRNA-Ile(GAU). Thus we see no reason why a reversal of this change should occur. Telford *et al.* (2000) have also discussed these two alternative scenarios in a parsimony treatment, however they did not consider the position of the tRNA-Ile(CAU) gene loss, or the fact that there are several independent modifications to the tRNA-Met, both of which are essential points in our interpretation.

Reassignment of AUA from Ile to Met has also been observed (Ehara *et al.* 1997) in some species of yellow-green algae (Xanthophytes). The entire genome of these species is not yet available and the lack of information about the tRNA genes and codon usage makes it impossible to deduce the mechanism of change. However, it is clear that the change occurred independently of similar changes observed in Metazoa and Fungi.

<u>Reassignments involving the AGR block are initiated by loss of a tRNA-Arg gene</u>

The AGR block, assigned to Arg in the canonical code, has undergone multiple reassignments in Metazoa. The tRNA-Arg(UCU) is present in Porifera but is absent in all other Metazoa. In Cnidaria, almost all tRNAs are missing, and tRNA import from the cytoplasm is required. This is not associated with codon reassignment. In the remaining Metazoa (*i.e.* Bilateria) the loss of the tRNA-Arg(UCU) gives rise to subsequent changes in the genetic code. Loss of this gene leaves the tRNA-Ser(GCU) as the only candidate for translating these codons. Hence, AGR is reassigned to Ser at the base of the Bilateria (Fig. 4). This is classed as the UC mechanism, as it is initiated by a tRNA loss. Note that AGR codons are frequent in Porifera and Cnidaria; therefore there is no suggestion of codon disappearance (Table 4).

The GCU anticodon of tRNA-Ser usually interacts with only AGY codons prior to the loss of the tRNA-Arg. However, pairing of the tRNA-Ser with AGR codons is also possible to some extent, once there is no longer any competition from the tRNA-Arg (Yokobori *et al.* 2001; Yokoyama et.al. 1995). In *Drosophila* it is known that the tRNA-Ser(GCU) translates AGA (Tomita et.al. 1999), *i.e.* there is a GA mispairing at the wobble position in the same way as in the mispairing of AUA with tRNA-Ile(GAU) after the loss of the tRNA-Ile(CAU). In *Drosophila,* AGG is absent, which may indicate that this codon is avoided because of selection against an unfavourable GG mismatch that would occur at the wobble position. It should also be noted that %G and %C are very low in *Drosophila* (Table 4), so the absence of AGG may simply be a result of mutation pressure. In *Daphnia* there is only 1 AGG codon, even though %G is high. This suggests definite avoidance of this codon. However, not all arthropods show this pattern. As an example, there are 12 AGG codons remaining in *Limulus*. Two other groups in which AGG is absent or very rare are the hemichordates and cephalochordates. Here the tRNA also has a G at the wobble position, which we presume is unmodified.

In contrast, Table 4 shows many other cases where AGA and AGG are both frequent. In several invertebrate groups, the G wobble position is mutated to a U, which allows it to pair with all the AGN codons. This is the case with *C. elegans* and *T. spiralis* in Table 4, where we see relatively high usage of AGG. Alternatively, in some Echinoderms (Matsuyama *et al.* 1998) and Molluscs (Tomita *et al.* 1998) a base modification from G to $m^7G$ also permits it to decode the entire AGN block as Ser.

The AGR block has undergone further changes in urochordates, where it is reassigned to Gly, and craniates, where it is reassigned to Stop. The most likely state prior to these changes would be that AGA was moderately frequent and AGG was rare or absent (as with the current hemichordates and cephalochordates). Urochordates contain an additional tRNA-Gly with anticodon UCU (Gissi *et al.* 2004). This has arisen by duplication of the standard tRNA-Gly(UCC), followed by anticodon mutation. This new tRNA pairs efficiently with AGR codons and outcompetes the tRNA-Ser(GCU). AGR codons become very frequent after the reassignment (Table 4) because mutation from GGR glycine codons occurs. Note that %A is greater than %G in all these genomes, so mutation pressure favours synonymous mutations from GGR to AGR. Since AGA was a moderately frequent codon for Ser prior to the reassignment, and since %A is high, we cannot argue that mutation pressure caused the disappearance of AGA and replacement by AGR prior to the change. Thus, it seems unlikely that codon disappearance is responsible for this change. This argument also applies for the reassignment of AGR to Stop in the craniates. AGA is rare in craniates after the change because Stop codons are always rare. These genomes have very high %A, so unless there was a short period when mutation pressure was in the opposite direction to that seen in all the current genomes, then we must suppose that this reassignment occurred without disappearance of the codon. This is puzzling since it implies premature termination of translation of genes containing AGA Ser codons.

Spruyt *et al.* (1998) argued that AGA is also read as Gly in *Branchiostoma lanceolatum*. They found a putative tRNA with anticodon UCU that may have arisen by duplication and mutation of the tRNA-Gly(UCC), as occurred in the urochordates. However, the codon usage numbers suggest that this is not the case, and that the sequence identified does not function as a tRNA-Gly. The pattern seen for AGA and AGG in *B. lanceolatum* resembles that in the hemichordates and some of the arthropods, where there is a single tRNA-Ser(GCU). Also, other cephalochordate genomes are now available, including *B. floridae* (Boore *et al.* 1999) and *Epigonichthys maldivensis* (Nohara *et al.* 2005), which also show no evidence of reassignment of AGA to Gly. The distinction between *B. lanceolatum* and *B. floridae* (e.g. Knight *et al.* 2001a) therefore seems unfounded.

<u>Reassignment of AAA from Lys to Asn may proceed via an Ambiguous Intermediate</u>

The reassignment of AAA from Lys to Asn occurred independently in Echinoderms and Platyhelminthes. Furthermore, the AAA codon is absent in the hemichordate *Balanoglossus* (see Table 4), but present in the related species, *Saccoglossus*, where it is translated as Lys, as in the canonical code (M. J. Smith, personal communication).

Most metazoans in which AAA retains its standard assignment of Lys have one tRNA-Lys(UUU), which pairs with AAR codons, and one tRNA-Asn(GUU), which pairs with AAY codons. Mispairing between a G in the anticodon wobble position and an A-ending codon is implicated in both the reassignment of AUA from Ile to Met and the reassignment of AGR from Arg to Ser, as discussed above. In the case of the tRNA-Asn(GUU), mispairing with AAA is prevented by post-transcriptional modification of the wobble position G to Queuosine (Q), which allows pairing with U and C but inhibits pairing with A (Morris *et al.* 1999).

In echinoderms and platyhelminthes, the tRNA-Lys anticodon is mutated to CUU (Tomita *et al.* 1999a), which pairs only with AAG. Several changes to the tRNA-Asn occur that increase its ability to pair with AAA. The modification of G to Q does not take place (Yokobori *et al.* 2001), which may allow some degree of GA mispairing. In echinoderms, modification of the second anti-codon position from U to Ψ (Pseudouridine) enhances the ability of the GΨU-Asn anticodon to recognize the AAA codon. Also in some echinoderms, the mutation of the base immediately adjacent to the 5' end of the anticodon from a U to a C seems to help the GUU anticodon to recognize AAA (Castresana *et al.* 1998). In *Balanoglossus* the tRNA-Lys has also changed to CUU, but there seem to be no modifications in the tRNA-Asn that would allow AAA to be reassigned to Asn. AAA is therefore unassigned.

In all groups in which AAA is still Lys, it is a relatively frequent codon and it is usually more frequent than AAG. %A is usually quite high at FFD sites. Thus, even though Castresana *et al.* (1998) argue for the CD mechanism, it seems unlikely to us that AAA disappeared in either the ancestor of the platyhelminthes or the echinoderms. Note that the absence of AAA in *Balanoglossus* is not shared with *Saccoglossus*, so we cannot use this to argue that AAA disappeared in the common ancestor of echinoderms and hemichordates. The AI mechanism seems the most likely mechanism for the reassignments in platyhelminthes and echinoderms. The process would have begun by a gain in function of the tRNA-Asn that gave it some ability to pair with AAA, and hence made AAA ambiguous, *e.g.* the Q modification ceased to occur. Then further changes to the tRNA-Asn would have increased the fraction of AAA codons translated as Asn, and a straightforward mutation of the tRNA-Lys could then have removed this ambiguity.

Alternatively, it is possible that the changes occurred via the UC mechanism. If the tRNA-Lys mutation happened first, this would leave the tRNA-Asn able to pair inefficiently with the AAA, and subsequent changes to the tRNA-Asn would allow it to recognize AAA more easily. This argument is analogous to the case of the AUA:Ile to Met change or the AGR:Arg to Ser. However, in those two cases the loss of function is the deletion of the original tRNA for the codon in question, which is irreversible. In the AAA case, the loss of function is just the mutation of U to C in the anticodon. This would be a deleterious mutation that could easily reverse, so it is difficult to see why the change would go to fixation. This makes the UC mechanism seem less plausible for AAA. We conclude, therefore, that of all the reassignments considered in this paper, these two examples of AAA reassignment are the best candidates for the AI mechanism.

In the case of *Balanoglossus*, the AAA codon cannot disappear by mutation pressure, and therefore we require a selective reason why this codon should be absent. This can be attributed to either UC or AI mechanisms. If a gain of function of the tRNA-Asn(QUU) brought about by the loss of Q modification happened first, the resulting tRNA-Asn(GUU) would have acquired some ability to decode AAA as Asn and consequently AAA would be ambiguously translated. The negative selection against such ambiguous translation may have driven the AAA codon to disappear. Subsequently, the loss of function of tRNA-Lys(UUU) occurred, brought about by a mutation from U to C at the first anticodon position, and this removed the ambiguity in AAA decoding. Further experimental information on the state of modification of the tRNA-Asn in *Balanoglossus* would be useful to confirm this.

Alternatively, the loss of function of tRNA-Lys, that prevents it translating AAA, may have occurred first. In the gain-loss framework, this would be a loss without a gain. AAA would be unassigned, and selection against this codon would drive its usage down to zero. This is analogous to the situation in *Schizosaccharomyces octosporus*, where we argued that the AUA codon was driven to zero after the loss of the tRNA-Ile(CAU) gene. However, this case is again made less plausible by the fact that the change in the tRNA-Lys would be reversible, whereas the loss of the gene from *S. octosporus* is irreversible. Indeed, the current situation in *Balanoglossus* seems unstable. A mutation of the tRNA-Lys(CUU) back to UUU would be neutral in the absence of AAA codons. If this occurred, then synonymous mutations of AAG codons to AAA would rapidly occur (since %A is greater than %G). This would drive the system back to the canonical code. The fact that AAA remains absent in *Balanoglossus* suggests that there is some reason why the reversal of the tRNA-Lys back to UUU would not be neutral. One possibility may be that CUU pairs more efficiently than UUU with the AAG codons, so translational efficiency would favour retention of the CUU anticodon once AAG became the dominant codon for Lys. Another possibility is that unmodified wobble position U bases can often pair with all four bases in four codon families. For this reason, wobble position U's are often modified when they occur in tRNAs for two-codon families only (Yokobori *et al.* 2001). Thus prior to the codon reassignment, the tRNA-Lys anticodon would be U*UU, where U* is a post-transcriptional modification of U that permits it to pair with A and G only. If the ability to modify the U were lost by *Balanoglossus* when the anticodon mutated to CUU, then a reversal of this mutation back to UUU would leave an unmodified U in the wobble position that could pair with both Asn and Lys codons. This would be selected against due to the ambiguity created in the Asn codons. Both these explanations are rather speculative.

<u>Introduction of new Stop Codons</u>

In addition to the AGR block, which becomes Stop in vertebrates, there are two further cases where a sense codon has been reassigned to Stop. According to the GenBank entry AF288091 (G. Burger, unpublished), UUA has been reassigned from Leu to Stop in the stramenopile *Thraustochytrium aureum* (Fig. 3). One tRNA-Leu gene in *T. aureum* has a CAA anticodon instead of the usual UAA anticodon, hence UUA is no longer translated as Leu. %A is high in this and related species, and UUA is frequently used as Leu in related species. This makes it unlikely that UUA disappeared due to mutation pressure. The mechanism cannot be assigned with certainty, but the most likely explanation is that UUA was driven to become rare after the mutation in the tRNA-Leu occurred because of inefficient pairing with the CAA anticodon. Gain of function of the release factor could then have occurred. This scenario corresponds to the UC mechanism (loss before gain). An alternative is that changes in the release factor began to occur in such a way that UUA became ambiguously translated as Leu and Stop. In this case, UUA would be selected against as a Leu codon and would be driven to low frequency, and selection would favour the mutation to C in the anticodon, so that UUA was no longer translated as Leu. This would correspond to the AI mechanism (gain before loss). It is difficult to decide between these scenarios.

A second similar case is the reassignment of UCA from Ser to Stop in *Scenedesmus obliquus* (Kück *et al.* 2000). In most species, a tRNA-Ser(UGA) would decode UCN codons, but in *S. obliquus*, the tRNA-Ser has a GGA anticodon that can only pair with the UCY codons. UCA has become the standard Stop codon, used in the 13 standard mitochondrial protein genes. Alongside this, UCG is absent, presumably because it no longer pairs with the tRNA-Ser. The Stop codons UAA and UGA are not used, while UAG has been reassigned to Leu (as discussed above under 'reassignments of UAG stop codons').

Table 5 shows that %A is fairly high in *S. obliquus* and related species, and that UCA is a fairly common codon in both *P. minor* and *C. eugametos*, where it retains its assignment to Ser (we will consider *C. reinhardtii* separately in the following section). It seems unlikely that UCA would disappear due to mutational pressure. Here again, it is possible to explain the result in terms of a UC or an AI scenario. In the UC scenario, the tRNA-Ser anticodon would mutate to GGA, which would drive UCA and UCG to very low values and permit the release factor to change so that it interacts with UCA. In the AI scenario, the release factor would change first, making UCA ambiguous and causing it to be selected against and driven to low frequency, and allowing the mutation in the tRNA-Ser to occur.

The full codon usage table of *S. obliquus* is given by (Kück *et al.* 2000, and Nedelcu *et al.* 2000). In Table 5 we have chosen selected codons in order to illustrate that *S. obliquus* is unusual in other ways. The usual Leu UUA codon is absent, and the tRNA-Leu has CAA anticodon, which translates UUG only (the same as in *T. aureum* except that UUA is not used as a Stop codon). AGA is also absent, but there is a tRNA-Arg with CCU anticodon that translates AGG. Nedelcu *et al.* (2000) identified a tRNA-Arg(UCU) that should interact with AGA, and a tRNA-Ile(UAU) that should interact with AUA. However, AGA and AUA are both absent. They state that these two tRNAs are redundant because their corresponding codons are not used. However, we argue that these two tRNAs must be non-functional or not expressed, otherwise the corresponding codons *would* be used. Note that %A > %G in *S. obliquus*, so if there were functional tRNAs with U at the wobble position, the A-ending codons would be frequent. As an illustration, the Glu codons in *S. obliquus* do follow the expected trend from the mutational frequencies: GAA is more frequent than GAG because these are both translated by a tRNA-Glu(UUC) with U at the wobble position. The Thr codons shown in Table 5 also have ACA more frequent than ACG, but one final peculiarity of the *S. obliquus* genome is that it possesses no tRNA-Thr. It is presumed that a tRNA-Thr(UGU) is imported from the cytoplasm.

**Importation of tRNAs from the cytoplasm to the mitochondria**

Although there are many mitochondrial genomes that possess sufficient tRNAs to translate the full genetic code, there are also many genomes in which this is not the case, and where import of nuclear-encoded tRNAs from the cytoplasm to the mitochondrion must be occurring. Table S1 in the supplementary information lists the amino acids for which tRNAs are present/absent in the mitochondrial genome of each species. In the figures, we have used a white arrow symbol to indicate positions where import of multiple tRNA genes must have evolved. We estimate at least 9 independent origins of tRNA import: *Acanthamoeba*, *Dictyostelium*, Cnidaria, Chytridiomycota, land plants, *Pedinomonas*, *Chlamydomonas*, *Naegleria*+Alveolata, and kinetoplastids (the latter are not shown but are known to use tRNA import). This count depends on the details of the phylogeny: it is possible that *Acanthamoeba* and *Dictyostelium* have a common ancestor, which would decrease the count by one, and it is possible that *Naegleria* is not a sister group to the alveolates, which would increase the count by one.

In all the above groups, at least 4 tRNAs are absent, and it is clear that the remaining tRNAs are insufficient, thus there is no doubt that import must occur. However, there are also many other groups where only a small number of tRNAs are missing. We have already discussed several cases above where the loss of a tRNA leads to the corresponding codon becoming unassigned or reassigned to a new amino acid. These cases leave clear signals in the codon usage patterns. This means that there is no import of a replacement tRNA from the cytoplasm in the cases discussed in previous sections. In contrast, we will now discuss several cases where a small number of tRNAs are absent from the mitochondrial genomes but there is no change in the genetic code or unusual codon usage pattern. In these cases it appears that import of one or a few specific tRNAs is occurring.

There is no tRNA for Thr in any of the stramenopile genomes. In addition, tRNA-Thr is absent from *Reclinomonas*, the Rhodophyta, *Mesostigma, Scenedesmus*, and all the groups where multiple tRNAs are missing. This makes tRNA-Thr the most frequently deleted tRNA. We consider this significant, but we have no plausible explanation why. The tRNA-Thr attracts attention because there are many genomes in which this is the only tRNA missing. Thr codons are nevertheless frequent in these genomes, so there must be a way of translating them. Import of tRNA-Thr from the cytoplasm is quite possible, but we do not know why this specific tRNA should be imported. Alternatively, in the genomes where tRNA-Thr is one of the only ones missing, it is possible that another tRNA is post-transcriptionally edited to become a tRNA-Thr. Two types of tRNA would thus arise from the same gene. This point has been suggested previously (Saks *et al.* 1998), but as far as we are aware, there is no direct evidence that this occurs.

There is no tRNA for the Arg CGN block in several of the stramenopiles (*Cafeteria, Thraustochytrium, Chrysodidymus, Ochromonas, Laminaria, Pylaiella*) but it is present in *Phytophthora* and *Saprolegnia*. Complete genomes are not available for the remainder of the stramenopiles shown on Figure 3; hence we cannot tell if the gene is present. If the phylogeny is correct, this implies three independent losses of the tRNA-Arg. Either tRNA import or some other tRNA change must be compensating. Recall that this same tRNA-Arg is also absent in several fungi and that this leads to the CGN block becoming unassigned (as shown of Fig. 1). So tRNA-Arg is not imported in the fungi.

Table S1 lists several other genomes where just one or two tRNAs are missing. It is possible that tRNA import occurs, or that there are unknown tRNA modifications that compensate for these individual losses. However, we consider this as somewhat uncertain in cases where only one genome is known with the missing gene. This could simply be a failure to locate the gene on the genome, or a problem of misannotation.

The codon usage of *Chlamydomonas reinhardtii* is markedly different from *C. eugametos* (Table 5 and Denovan-Wright *et al.* 1998). Both of these species have only three tRNAs (Table S1, supplementary information), hence tRNA import is required. *C. eugametos* uses almost all codons, and the codon usage is in line with what we would expect from the FFD site frequencies. However, there are many codons that are absent entirely from the *C. reinhardtii* genome. This suggests that a restricted set of tRNAs is being imported in *C. reinhardtii*, which causes selection against codons that cannot be translated, whereas a much broader set is being imported in *C. eugametos*, which allows a more standard pattern of codon usage. There are several parallels between the codon usage in *C. reinhardtii* and *Scenedesmus obliquus* (a sister species, see Fig. 2). UUA (Leu), AUA (Ile), UCG (Ser) and AGA (Arg) are all absent from both. In *S. obliquus,* this is due to peculiarities in the tRNAs encoded by the mitochondrial genome (see previous section). The correspondence between the codon usage and tRNA set in *S. obliquus* shows that there is no tRNA import in this species. It therefore appears that the tRNAs that were first imported in *C. reinhardtii* correspond to those that were needed according to the codon usage pattern of the ancestor of *C. reinhardtii* and *S. obliquus*. In *C. eugametos*, the evolution of tRNA

import has proceeded further. A more complete set is imported and the codon usage has relaxed back to a more standard pattern due to mutation.

## Discussion

Distinguishing the mechanisms

The gain-loss framework that we use to classify mechanisms of codon reassignment was introduced in discussion of our computer simulations (Sengupta and Higgs, 2005). Here we wish consider to what extent this classification is applicable to the real events. The CD mechanism is defined by the fact that the codon disappears from the genome and is replaced by synonymous codons prior to any changes in the tRNAs and release factors. The reasons for the disappearance are random drift and directional mutation pressure. In simulations it is obvious whether a codon has disappeared. In real examples, there is clear evidence if the codon usage is zero or close to zero in the species close to the reassignment point, and there is also clear evidence if mutation pressure is in the right direction to cause codon disappearance. It is, of course, never possible to say that the codon usage was exactly zero in some genome in the past. A practical definition of the CD mechanism for real cases is that absence or extreme rarity of the codon is the major factor that initiates the codon reassignment. There should be evidence that the codon was rare and that the probability of codon disappearance was not too small. We have shown that this is true in the cases where we argue that the CD mechanism occurred. Therefore, the distinction of CD from the alternatives is one that we feel can be reliably made in real examples.

In cases where we have determined that the codon did not disappear, the central question is to distinguish between UC and AI mechanisms. The UC mechanism is defined by the fact that the loss occurs before the gain, whereas in the AI mechanism, the gain occurs before the loss. Although this distinction is obvious in simulations, it can be more difficult in real cases, as we did not observe the order of events. Nevertheless, we gave several examples where we feel a reliable classification of UC can be made. In particular, the reassignments of AUA from Ile to Met and AGR from Arg to Ser are both associated with the deletion of a tRNA from the genome. The loss event is an irreversible gene deletion that leaves the organism in a deleterious state. The gain event is then positively selected in response to this loss. In contrast, in the case of the reassignment of AAA from Lys to Asn, the loss event is a mutation in the anticodon of the tRNA-Lys. This is reversible, so selection would favour mutation back to the original state, which seems more likely to occur than making a codon reassignment. The gain in this case is probably due to the cessation of the Q modification process, which could occur due to deleterious mutations in the modifying enzyme or disruption of the transport of the enzyme to the mitochondria. These changes seem less easily reversible than a single mutation in the anticodon. Therefore we argue that the gain occurred first and that the loss occurred in response. This makes the AAA reassignment the most likely example of the AI mechanism. The cases where both loss and gain seem easily reversible are the most difficult to classify, and both AI and UC scenarios can be proposed. (These arguments are based on the assumption that the ambiguous and unassigned states are deleterious with respect to the original code. We deal with the alternative suggestion that ambiguous translation can be positively selected later in the discussion.)

We just argued that the AUA Ile to Met case is initiated by deletion of the tRNA-Ile(CAU) and that this change can be reliably classed as UC. Nevertheless, it is possible that ambiguous translation plays a role in this reassignment in the following sense. After the deletion, the AUA codon would be translated inefficiently by the tRNA-Ile(GAU), but it is possible that there might also be some interaction with the tRNA-Met(CAU). When the tRNA-Met becomes modified, it definitely gains the ability to translate AUA. So, ambiguous translation of AUA as both Ile and Met at some points in this process is not unlikely. Before the original tRNA-Ile(CAU) is deleted, any slight ability of the other two tRNAs to translate AUA would be irrelevant. Also, after the modification of the tRNA-Met, any slight ability of the tRNA-Ile(GAU) to translate AUA would be irrelevant. The ambiguity, if it existed, would only be relevant when there is a competition between two poorly adapted tRNAs. Despite all this, the gain in function of the tRNA-Met only occurs after the loss of the tRNA-Ile(CAU), so this reassignment counts as UC not AI. We recommend the use of the term AI only when ambiguous translation occurs as a result of a gain occurring before a loss, and where competition occurs between two well-adapted tRNAs.

The reassignment of AGR from Ser to Gly in urochordates involves a gain due to the duplication and anticodon mutation of the tRNA-Gly. In this case too, it is plausible that an AI stage for AGR decoding may have resulted after the gain of the new tRNA-Gly. The ambiguity would eventually be removed due to the selection of the superior recognition of AGR by the new tRNA-Gly. However, this only happens *after* the deletion of the original tRNA-Arg(UCU). We have counted the changes from Arg to Ser and Ser to Gly as separate events because the amino acid changes twice. However, we may think of the duplication of the tRNA-Gly as the gain that finally compensates for the loss of the tRNA-Arg considerably earlier in evolution. The AGR:Ser to Gly case is one of the examples where the number of tRNAs increases (see Table 6 and discussion below), but this only returns the number back to its value in the canonical code; it does not represent an increase with respect to the canonical code.

A slow response to a tRNA deletion occurs both for the tRNA-Ile(CAU) and the tRNA-Arg(UCU), which are deleted at roughly the same time in the ancestral Bilateria. This would leave AUA inefficiently translated as Ile and AGR inefficiently translated as Ser. Both these situations remain in some phyla today. However, in other groups, AUA has been captured by Met, and AGR has either become a useful Ser codon (due to mutation of the tRNA-Ser) or has been reassigned to Gly or Stop. These secondary changes may have occurred considerably later than the deletions of the original two tRNAs.

The fourth mechanism occurring in the gain-loss framework is the Compensatory Change mechanism. This would correspond to a case where the loss and gain become fixed in the population at the same time. Organisms with either the gain or loss but not both would only occur as rare variants, and would never be frequent in the population. This mechanism is expected in the theory due to the analogy of the problem with compensatory mutation theory (Kimura, 1985; Higgs, 1998, 2000). In the simulations of codon reassignment (Sengupta & Higgs, 2005) it was possible to distinguish a few cases where this occurred. It may well be that the CC mechanism occurs in real cases, and it could be an explanation in cases where both AI and UC seem equally plausible. However, it would be very difficult to distinguish this from either the UC or AI mechanisms after the event.

Comparison with previous surveys

Recent papers by Knight *et al.* (2001b) and Swire *et al.* (2005) have also carried out surveys of changes in the mitochondrial genetic code. Knight *et al.* (2001b) highlighted the fact that variation in GC content plays a major role in the CD mechanism. Most mitochondrial genomes are subject to mutation pressure from GC to AU. Since third codon positions are often synonymous, they argued that the CD mechanism would predict that the codons that are reassigned should have C or G at the third position. As this is not always the case, they argued against the CD mechanism. However, it is not the third codon position *per se* that matters; it is the position at which the synonymous mutation occurs. Table 6 gives examples where synonymous mutations driven by GC to AU mutation pressure have occurred at all three positions. Knight *et al.* (2001b) came down strongly in favour of the AI mechanism. We wish to redress the balance here. Our data show the CD mechanism is the best explanation in over half the cases. We agree with Swire *et al.* (2005) that CD applies well to stop codon reassignments, but does not usually apply to sense codon reassignments. The cases of CUN and CGN reassignments are exceptions to this, where we argue for CD with sense codons.

In their proposal of the AI mechanism, Schultz and Yarus (1994) identified three specific types of codon-anticodon mispairing that might be important in allowing ambiguous translation of a codon: GA at 3rd codon position; CA at 3rd codon position; and GU at 1st codon position. Knight *et al.* (2001b) considered whether these mispairing events were implicated in the reassignments in mitochondrial genomes. In Table 6, we also do this (*c.f.* Table 3 of Knight *et al.*). We agree that GA mispairing at the 3rd codon position is important in the reassignments of AAA and AGR. For the AUA Ile to Met transition, we argue that GA mispairing between tRNA-Ile(GAU) and the AUA codon is important, otherwise deletion of the tRNA-Ile(CAU) would not be possible. Mispairing of the unmodified tRNA-Met(CAU) with AUA (as in the Knight *et al.* table) might also be relevant, but we do not think this is significant prior to the deletion of the tRNA-Ile(CAU). The reverse transition of AUA from Met back to Ile listed by Knight *et al.* does not occur in our interpretation (Figure 4). Also the reassignment of UAA from Stop to Tyr listed by Knight *et al.* is no longer thought to occur. Knight *et al.* argue that CA mispairing is important for the UGA Stop to Trp transition. There is evidence for this in *Bacillus subtilis* (Lovett *et al.* 1991, Matsugi *et al.* 1998), where UGA is ambiguous. It is not clear whether this is a key feature in the UGA reassignments in mitochondria. There are many species where the canonical system works fine (UGA is Stop, UGG is Trp, and the tRNA-Trp has CCA anticodon). It is possible that if CA mispairing started to occur due to some change in the tRNA-Trp outside the anticodon, UGA would become ambiguous, and this would provide a selective pressure against UGA codons that would help to drive them to disappear. However, we are not aware of any evidence for this in mitochondria. The disappearance of UGA is likely to occur in any case due to mutation pressure and drift, and this seems to be a more important factor. The other cases listed by Knight *et al.* where mispairing is implicated do not occur in mitochondria, and we did not consider them. From Table 6, we would argue that mispairing is important in some but not all reassignments. However, the existence of mispairing does not necessarily imply the AI mechanism occurred, and mispairing is not a diagnostic feature of AI, as previously envisioned.

If translation of a codon becomes inefficient due to tRNA loss, then selection will act against it and it will tend to be replaced by synonymous codons. If this situation remains for some time without any compensating gain of function in another tRNA, then the codon may in fact disappear. We have argued that this is the case in *S. octosporus*, where the tRNA-Ile(CAU) has been deleted and AUA has disappeared, and also in the two *Kluyveromyces* species, in which AUA is very rare (Table 3). A similar thing happens with the AGR reassignment: there are several invertebrate groups in which AGG becomes rare after the deletion of the tRNA-Arg because of inefficient pairing with the tRNA-Ser when it retains its original GCU anticodon (Table 4). In cases where there is a full gain in function of the new tRNA, the codon in no longer selected against. This is seen both when modifications occur in the tRNA-Met and the tRNA-Ser.

Given that the loss of the tRNA in the first step of the UC mechanism is likely to be deleterious, anything that offsets the deleterious effect to some extent will increase the likelihood of the UC mechanism occurring. It has been suggested that deletion of a tRNA gene might be favoured since it reduces the total length of the genome and hence allows more rapid replication (Andersson and Kurland 1991 and 1995). We considered this effect in our simulations (Sengupta and Higgs, 2005) and showed that selection for reduction in genome length does indeed increase the likelihood of the UC mechanism. This factor will be more important in smaller genomes because the relative change in length for deleting a tRNA will be larger. For the cases of deletion of the tRNA-Ile and tRNA-Arg that we are considering, we do not have an experimental measurement of the selective disadvantage to the translation system of loss of the gene, so it is not possible to say whether the potential advantage of shortening the genome is large or small compared to this. We do not wish to argue that these reassignments are 'caused' by selection to reduce the genome size, but we do wish to say that it is the chance deletion of these tRNAs that initiated the reassignment in these cases, and that the fixation in the population of the genome with the deletion could well be aided by selection for more rapid genome replication.

However, there is no general trend for tRNA deletion at codon reassignment. We have given the changes in tRNA number associated with our interpretation of the reassignment events in Table 6. There are three cases of tRNA gain, each of which occurs once. These are due to a tRNA duplication followed by an anticodon mutation. There are four cases of tRNA deletion, which together occur ten times. Changes in release factors are not included in this table because these genes are not coded by the mitochondrial genome. In Table 6 in the supplementary information of Swire *et al.* (2005), a distinction is made between the theoretical and observed changes in tRNA number. Given the genome information now available, this distinction is no longer necessary. The AGR Arg to Ser reassignment in Bilateria is classed as an observed gain by Swire *et al.* (their case 9) because they are comparing with Cnidaria. However, it is clear that tRNA import has evolved separately in Cnidaria, and that it does not occur in either Porifera or the ancestral Bilaterian. The proper comparison for Bilateria is with Porifera, where the tRNA-Arg is present; therefore it is evident that the gene has been deleted in Bilateria at the point of the codon reassignment. Case 6 of Swire *et al.* is the AUA Ile to Met that occurs in yeasts. We have shown in this paper that the tRNA-Ile(CAU) gene is present in the fungi prior to this reassignment and is deleted at the point of the reassignment. Once again, this is a clear deletion and there is no difference between the observed and the expected change.

An important issue in the AI mechanism is the nature of the selective effect on an ambiguous codon. In our model (Sengupta & Higgs, 2005), we assumed that ambiguous translation would be unfavourable because the wrong amino acid would be inserted part of the time. However, it has been suggested that ambiguous translation may be favourable (Santos *et al.* 1999). Swire *et al.* (2005) also gave an argument related to the cost of amino acid biosynthesis, whereby ambiguous translation might be favourable because the saving in cost of using an energetically cheaper amino acid might outweigh the penalty of the ambiguity. The ambiguity penalty may be very small at particular sites where the amino acid substitution is neutral or almost neutral; hence, the cost saving would make this change advantageous. This mechanism is possible and we are currently in the process of studying it using simulations of the same type as those in Sengupta & Higgs (2005). A problem we see with this argument is that, if an amino acid substitution were favoured by cost saving, it would be easier to achieve this by a straightforward mutation in the gene without changing the genetic code. Such mutations could occur in places where it was favourable and not in places where it was not. There must be some sites in the protein where the substitution is deleterious, even if there are also some neutral sites. On the other hand, if the codon were reassigned, this would force changes in all sites in the protein, whether favourable or not, which would act against the postulated benefit of the change.

We do not think that positive selection for ambiguous translation can be a general explanation of sense to sense reassignments. We have already given one case where a sense to sense reassignment is better explained by the CD mechanism and several where it is better explained by UC. Swire *et al.* (2005) did not distinguish between AI and UC. However, the cost-saving argument is less relevant for UC than AI. In AI, if the cost-saving scenario occurred, it would be central to the argument because it would provide a reason why the gain occurred first, *i.e.* it would be the driving force for the reassignment. In the UC mechanism, the cause of the transition is the chance loss event, which is presumed to be deleterious and is not driven by selection. The gain event would then be positively selected because it would get rid of the penalty from the slow translation that occurs when there is no well adapted tRNA. The cost saving argument might also favour the gain event in this case. This would change the details of the selective forces favouring the gain, but this would only be relevant after the initial deletion. However, there is no need to postulate the existence of the neutral sites and cost-saving in the UC case, because the gain would be driven by selection to improve translational efficiency anyway. Thus, if the cost-saving scenario occurred in a UC reassignment, it would be a minor detail, not a driving force.

In any case, the best documented example of ambiguous translation of a codon is the CUG codon in the *Candida* nuclear genome (Santos *et al.* 1999; 2004), and this example now provides evidence that ambiguous translation is deleterious. It has been shown that the reassignment is driven by the gain of a new tRNA-Ser with anticodon CAG that translates the CUG codon, which is normally Leu. This case was formerly cited as an example of positive selection for ambiguous translation. However, Massey *et al.* (2006) have now shown that there is selection against the ambiguous CUG codons that causes them to become very rare in their original Leu positions. Most occurrences of CUG in *Candida* are newly evolved in Ser positions. This shows that there is negative selection on the ambiguous codon. (Note that this is the third reason for a codon to become rare: in the CD mechanism, the codon disappears due to mutation pressure; in the UC mechanism, the codon is inefficiently translated and selected against and therefore might become rare; in the AI mechanism, the codon might be selected against because it is ambiguous, and therefore might become rare.) Another study suggesting ambiguous translation is deleterious is that of Kim *et al.* (2000), who studied many examples of anticodon mutations in tRNA-Phe from *E. coli*. In cases where the cells expressing the mutant tRNA were viable, it was found that the mutant tRNA was charged with the amino acid corresponding to the new anticodon, *i.e.* there was no ambiguous translation. However, in other cases, expression of the mutant tRNA was lethal. It was presumed that the mutant tRNA was still charged with Phe although its anticodon corresponded to another amino acid. This would cause ambiguous translation, which was apparently deleterious in the experiment.

The arguments in this paper apply only to codon reassignments that have occurred since the establishment of the canonical code. They do not apply to the origin and build up of the canonical code. We believe that there is a good case that positive selection was important in the establishment of the canonical code (Freeland *et al.* 2003; Ardell & Sella, 2002) and that this accounts for the non-randomness and apparent optimization of the code. Adding a new amino acid to the code opens up a whole new realm of protein sequences that can be formed using the new amino acid. Therefore, there is ample scope for positive selection to act in determining the codons to which each amino acid is initially assigned. Translational error rates and mutation rates may have been large in early organisms, so minimizing the effects of these errors may have been important. However, we do not think that positive selection guides the codon reassignments that have occurred after the establishment of the canonical code. There is little to be gained from a codon reassignment in modern organisms because no new amino acid is being added, and the effects of minimizing error will be very small when both translational error and mutation rates are as small as they are today.

tRNA evolution

Although Table 6 shows no strong trend to reduce the number of tRNAs at codon reassignment events, we feel that this obscures the very large role that tRNA deletion has had in the evolution of the mitochondrial translation system. A major difference between bacterial and mitochondrial systems is that four-codon families in bacteria require at least two tRNAs - one with wobble position G to translate U- and C-ending codons, and one with wobble position U to translate A- and G-ending codons. Most mitochondria require only one tRNA with wobble position U for all four codons. In metazoan mitochondria the second tRNA has been deleted in all eight four-codon families. The same genes have been deleted in other groups too, and we suspect that there have been independent deletions of the same gene, although demonstrating this would be difficult because we do not have a fully resolved tree for the earliest branches of the eukaryotes. Nevertheless, the same process has occurred at least once in all eight four-codon families, which is relevant to the discussion on genome streamlining, even though there is no change in the code. Loss of a tRNA with wobble position G may have been slightly deleterious originally when it occurred in the early stages of mitochondrial evolution, but it is possible that the ribosome has become sufficiently flexible to tolerate this. It could also be that the loss still comes with a price of decreased accuracy and/or speed of translation. Selection for reduction in genome length might well play a role in these deletions also. The 22 tRNAs in animal mitochondria represent the minimal possible set. It is only possible to reduce the number further by making codons unassigned or by evolving a mechanism of tRNA import from the cytoplasm. We have seen that tRNA import has evolved multiple times. If import is possible, then there will be no deleterious effect of further tRNA deletions from the mitochondrial genome, and selection for shorter genome length may favour the fixation of variants in which tRNAs have been deleted. Thus rapid tRNA deletion might be expected in lineages in which a general import process is operating.

It is known that many protein genes formerly present on the mitochondrial genome have been transferred to the nuclear genome (Blanchard and Lynch, 2000). Many of these code for proteins that are required in the mitochondrion and are targeted back to the organelle after synthesis. In cases where tRNAs are imported, it is interesting to ask whether these tRNAs are in fact formerly mitochondrial tRNA genes that have been transferred to the nucleus, or whether the existing nuclear tRNAs have become functional in the mitochondria and the mitochondrial genes have simply been deleted. We attempted to answer this question with the kinetoplastids, *Trypanosoma brucei* and *Leishmania tarentolae*, which possess no mitochondrial tRNAs and where import is known to occur (Simpson et.al. 1989; Hancock and Hajduck, 1990; Schneider and Marechal-Drouard, 2000). The full set of tRNAs in the nuclear genome of these species is known. We carried out phylogenetic analysis comparing these genes with tRNAs from other eukaryotic nuclei, bacteria and mitochondria. For each separate tRNA, the kinetoplastid genes clustered with eukaryotic nuclear genes and not with mitochondria. We do not show these results because the trees have very low statistical support due to the short sequence length of the tRNAs. Nevertheless, we found no evidence that any of the kinetoplastid tRNAs had been transferred from the mitochondrial genome.

The identity elements and anti-determinants of tRNAs are key aspects of the molecules that play a crucial role in the proper aminoacylation and prevention of mis-acylation of a tRNA (Giegé *et al.* 1998). These are important factors in understanding changes in the genetic code. Most aminoacyl tRNA-synthetases use the anticodon as a key identity element for proper recognition of the tRNA. Two exceptions are Leu and Ala where the anticodon has no role to play in the proper charging of the cognate tRNA with the amino-acid. In such cases, mutations in the anticodon do not affect the aminoacylation of the tRNA. Such mutations can result in codon reassignment if the resulting anticodon acquires the ability to pair with a codon originally associated with a different amino acid. The reassignment of UAG from Stop to Leu in the mitochondrial genome of some species of green plants provides such an example (Laforest *et al.* 1997). Another possibility is that mis-acylation of a tRNA due to a change in an aminoacyl tRNA-synthetase could be a cause of a codon becoming ambiguous. The tRNA would be ambiguously charged, but there would only be one type of tRNA for the codon. However, this does not seem to occur in the mitochondrial reassignments we considered here.

## Conclusions

We have given arguments above as to which mechanisms seem most likely in each of the codon reassignments in mitochondrial genomes. We have shown that the many reassignments of stop codons to sense codons are readily explained by codon disappearance, given the biased base composition of most mitochondrial genomes and the small total number of occurrences of stop codons in these genomes. Disappearance of sense codons is more difficult because the total number of codons for the corresponding amino acid is large. A very strong mutational bias is required for sense codons to disappear. However, in one group of yeast species, an extreme bias against C does exist, and we argue that the reassignment of CUN and CGN codons in these species is attributable to codon disappearance. In the other examples of sense codon reassignments, the mutational bias is in the wrong direction for causing codon disappearance, and the probability of disappearance is negligible.

Where the codon does not disappear, we have emphasized the important distinction between the UC and AI mechanisms. The case for the UC mechanism is most clear when the reassignment is associated with a tRNA deletion. We then argue that the deletion initiated the process and the codon reassignment occurred as a response to this. The UC mechanism does not rely on selection for reducing genome length, but if such selection were significant, this would increase the likelihood of this mechanism. Many non-essential tRNAs have been deleted during mitochondrial genome evolution and these did not initiate codon reassignments because the original code was still functional after the deletion. However, this makes it clear that chance tRNA deletion is a relatively common event. We also observed several cases where a good argument can be made for the AI mechanism, *i.e.* where the reassignment arose because the codon first became ambiguous. In other cases, scenarios for both AI and UC seemed equally plausible, and it is difficult to distinguish them after the event.

We see these genetic code changes as chance events, rather than as changes governed by positive selection. Disappearance is a chance event that occurs under drift when there is strong mutational bias in base frequencies. If a change in tRNAs or release factors happens to occur while a codon is absent, then a codon reassignment can occur. However, this is a chance event: the codon frequency could drift back to a higher level without any reassignment occurring. Our interpretation of reassignments via UC and AI mechanisms is that they too are initiated by chance events, such as the deletion of a tRNA gene or the change in process of base modification in an anticodon. These changes are probably slightly deleterious, but efficient functioning of the translation system can be restored by making the codon reassignment. The view that ambiguous intermediate states are driven by positive selection seems unlikely to us at present. The origin of the canonical code is outside the scope of the present paper. However, we emphasize that the situation in codon reassignments in modern organisms is different from that during the early evolution of the canonical code, where positive selection probably had an important role.

We conclude that our gain-loss framework is suitable as a description of the real codon reassignment events. It emphasizes that there are several mechanisms that are alternatives within a larger picture, and that it is not always profitable to discuss these mechanisms as though they were mutually exclusive. These mechanisms can and do occur in nature, and one mechanism is not sufficient to explain all cases.

**Acknowledgements**
This work was supported by the Canada Research Chairs and the Natural Sciences and Engineering Research Council of Canada.

Table 1: Codon usage data relevant to reassignments of Stop codons UGA and UAG.
[a] These species have CCA anticodon but UGA is Trp. This suggests the C base is modified in the tRNA.
[b] UAG is reassigned to Leu in these species.
[c] In vertebrates AGA and AGG are also Stop codons. Each is used once in *Homo sapiens*.
[d] In *S. obliquus*, UCA is a novel Stop codon occurring 17 times.
[e] These codons are presumed to be inefficiently translated as Trp by the tRNA-Trp(CCA).

| | UGA is | tRNA-Trp anticodon | Codon usage UGA | UAA | UAG |
|---|---|---|---|---|---|
| <u>Species in Figure 1</u> | | | | | |
| *Amoebidium castellanii* | Trp | [a]CCA | 78 | 28 | 12 |
| *Dictyostelium discoideum* | Stop | CCA | 2 | 30 | 8 |
| *Monosiga brevicolis* | Trp | TCA | 124 | 28 | 4 |
| *Homo sapiens*[c] | Trp | TCA | 92 | 8 | 3 |
| *Rhizophidium sp.* 136 | Stop | CCA | 5 | 32 | 117[b] |
| *Spizellomyces punctatus* | Stop | CCA | 12 | 20 | 140[b] |
| *Hyaloraphidium curvatum* | Not used | CCA | 0 | 16 | 2 |
| *Monoblepharella sp. JEL 15* | Stop | CCA | 1 | 20 | 5 |
| *Harpochytrium sp. JEL105* | Not used | CCA | 0 | 12 | 2 |
| *Harpochytrium sp. JEL94* | Not used | CCA | 0 | 13 | 1 |
| *Allomyces macrogynus* | Not used | CCA | 0 | 16 | 14 |
| *Mortierella verticillata* | Rare | CCA | 2[e] | 21 | 4 |
| *Rhizopus oryzae* | Not used | CCA | 0 | 20 | 4 |
| *Crinipellis perniciosa* | Trp | [a]CCA | 124 | 73 | 16 |
| *Schizophyllum commune* | Trp | [a]CCA | 25 | 19 | 1 |
| *Penicillium marneffei* | Trp | TCA | 61 | 16 | 1 |
| *Hypocrea jecorina* | Trp | TCA | 89 | 14 | 5 |
| *Schizosaccharomyces japonicus* | Not used | CCA | 0 | 6 | 1 |
| *Schizosaccharomyces octosporus* | Not used | CCA | 0 | 8 | 0 |
| *Schizosaccharomyces pombe* | Rare | CCA | 1[e] | 7 | 0 |
| *Yarrowia lipolytica* | Trp | TCA | 57 | 12 | 2 |
| *Candida stellata* | Trp | TCA | 34 | 8 | 0 |
| *Candida albicans* | Trp | TCA | 49 | 5 | 8 |
| *Saccharomyces cerevisiae* | Trp | TCA | 124 | 19 | 0 |
| | | | | | |
| <u>Species in Figure 2</u> | | | | | |
| *Malawimonas jakobiformis* | Stop | CCA | 1 | 46 | 2 |
| *Cyanidioschyzon merolae* | Stop | CCA | 2 | 29 | 3 |
| *Chondrus crispus* | Trp | TCA | 101 | 22 | 3 |
| *Porphyra purpurea* | Trp | TCA | 117 | 27 | 4 |
| *Chaetosphaeridium globosum* | Stop | CCA | 7 | 30 | 9 |
| *Chara vulgaris* | Stop | CCA | 8 | 30 | 8 |
| *Prototheca wickerhamii* | Not used | CCA | 0 | 35 | 1 |
| *Pseudoendoclonium akinetum* | Stop | CCA | 11 | 40 | 21 |
| *Pedinomonas minor* | Trp | TCA | 62 | 11 | 0 |
| *Sscenedesmus obliquus*[(d)] | Stop | CCA | 1 | 2 | 141[(b)] |
| *Chlamydomonas eugametos* | Not used | CCA | 0 | 12 | 2 |
| *Chlamydomonas reinhardtii* | Not used | CCA | 0 | 6 | 2 |
| | | | | | |
| <u>Species in Figure 3</u> | | | | | |
| *Emiliana huxleyi* | Trp | UCA | 73 | 19 | 2 |
| *Rhodomonas salina* | Stop | CCA | 1 | 34 | 9 |
| *Naegleria gruberi* | Not used | CCA | 0 | 37 | 9 |
| *Plasmodium reichenowi* | Not used | none | 0 | 3 | 0 |
| *Plasmodium falciparum* | Not used | none | 0 | 3 | 0 |
| *Paramecium aurelia* | Trp | TCA | 83 | 29 | 17 |
| *Tetrahymena pyriformis* | Trp | TCA | 228 | 44 | 0 |
| *Tetrahymena thermophila* | Trp | TCA | 228 | 45 | 0 |
| *Caferteria roenbergensis* | Trp | TCA | 190 | 32 | 2 |
| *Phytophthora infestans* | Stop | CCA | 1 | 39 | 0 |
| *Saprolegnia ferax* | Not used | CCA | 0 | 42 | 1 |
| *Chrysodidymus synuroides* | Not used | CCA | 0 | 34 | 3 |
| *Ochramonas danica* | Not used | CCA | 0 | 30 | 14 |
| *Laminaria digitata* | Stop | CCA | 4 | 29 | 6 |
| *Pylaiella littoralis* | Stop | CCA | 7 | 38 | 7 |

Table 2

| Case | Species | $N_{stop}$ | %G | %A | $f_{codon}$ (%) | $P_{dis}$ |
|---|---|---|---|---|---|---|
| UGA codon disappearance | | | | | | |
| (i) | *Monosiga brevicolis* | 32 | 2.85 | 47.52 | 5.36 | 0.17 |
| (ii) | *Acanthamoeba castellanii* | 40 | 9.89 | 26.35 | 21.44 | 6.4E-05 |
| (iii) | *Crinipellis perniciosa* | 89 | 9.19 | 32.84 | 17.94 | 2.3E-08 |
| (iii) | *Schizophyllum commune* | 20 | 3.48 | 42.87 | 6.99 | 0.23 |
| (iv) | *Penicillium marneffei* | 17 | 2.38 | 35.64 | 5.89 | 0.36 |
| (iv) | *Hypocrea jecorina* | 19 | 4.68 | 40.14 | 9.46 | 0.15 |
| (v) | *Yarrowia lipolytica* | 14 | 1.55 | 49.01 | 2.97 | 0.65 |
| (v) | *Candida stellata* | 8 | 0.94 | 41.57 | 2.17 | 0.84 |
| (vi) | *Chondrus crispus* | 25 | 6.08 | 36.61 | 12.46 | 3.6E-02 |
| (vi) | *Porphyra purpurea* | 31 | 10.98 | 35.80 | 19.01 | 1.4E-03 |
| (vii) | *Pedinomonas minor* | 11 | 1.60 | 21.31 | 6.52 | 0.48 |
| (viii) | *Emiliana huxleyi* | 21 | 9.26 | 36.50 | 16.83 | 2.1E-02 |
| (ix) | *Paramecium aurelia* | 46 | 14.87 | 19.60 | 30.14 | 6.8E-08 |
| (ix) | *Tetrahymena pyriformis* | 44 | 3.22 | 42.27 | 6.62 | 4.9E-02 |
| (ix) | *Tetrahymena thermophila* | 45 | 2.63 | 40.46 | 5.75 | 7.0E-02 |
| (x) | *Cafeteria roenbergensis* | 32 | 6.17 | 32.56 | 13.74 | 6.6E-03 |
| UAG codon disappearance | | | | | | |
| (i) | *Rhizophidium* | 37 | 3.33 | 34.05 | 8.18 | 4.3E-02 |
| (i) | *Spizellomyces punctatus* | 32 | 6.03 | 30.69 | 14.10 | 7.7E-03 |
| (ii) | *Scenedesmus obliquus* | 3 | 6.64 | 35.73 | 13.55 | 0.65 |

Table 3: Codon usage in some Fungi lineages.
(E) Intronic ORFs excluded; (I) Intronic ORFs included.
[a] CUN is reassigned to Thr and an unusual tRNA-Thr gene is present in these species.
[b] There is no tRNA for the CUN block in this species.
[c] The tRNA-Arg is deleted in these species.
[d] The tRNA-Ile(CAU) is deleted in these species.
[e] AUA is reassigned to Met in these species.

| | Leu codons | | Arg codons | | Ile and Met Codons | | | | Frequency at FFD sites | | | |
|---|---|---|---|---|---|---|---|---|---|---|---|---|
| | CUN | UUR | CGN | AGR | AUU | AUC | AUA | AUG | %U | %C | %A | %G |
| *S. japonicus (E)* | 79 | 198 | 7 | 32 | 133 | 40 | 32 | 48 | 76.0 | 3.4 | 19.2 | 1.3 |
| *S. octosporus (E)* | 68 | 236 | 2 | 34 | 161 | 34 | 0[d] | 57 | 61.1 | 1.9 | 34.8 | 2.2 |
| *S. pombe (E)* | 53 | 192 | 7 | 33 | 113 | 39 | 49 | 51 | 56.4 | 1.3 | 40.5 | 1.8 |
| *Y. lipolytica (E)* | 44 | 618 | 0[c] | 75 | 174 | 87 | 277 | 119 | 48.3 | 1.1 | 49.0 | 1.5 |
| *C. stellata (E)* | 3 | 279 | 12 | 29 | 123 | 8 | 156 | 54 | 57.3 | 0.2 | 41.6 | 0.9 |
| *C. albicans* | 132 | 397 | 47 | 26 | 119 | 81 | 229 | 100 | 55.4 | 4.8 | 37.7 | 2.1 |
| *C. parapsilosis (E)* | 66 | 547 | 39 | 45 | 303 | 32 | 193 | 117 | 68.6 | 0.7 | 29.8 | 0.9 |
| *C. parapsilosis (I)* | 137 | 728 | 60 | 102 | 410 | 49 | 299 | 143 | 65.3 | 3.0 | 29.0 | 2.7 |
| *P. canadensis (E)* | 25 | 714 | 18 | 67 | 274 | 18 | 562 | 105 | 49.2 | 0.7 | 49.4 | 0.7 |
| *P. canadensis (I)* | 27 | 746 | 20 | 74 | 298 | 20 | 586 | 109 | 50.0 | 0.9 | 48.4 | 0.6 |
| *A. gossypii* | 80[a] | 291 | 0[c] | 40 | 215 | 7 | 95[d,e] | 34 | 57.2 | 0.0 | 42.8 | 0.0 |
| *K. lactis (E)* | 0[b] | 286 | 0[c] | 48 | 213 | 16 | 7[d] | 63 | 44.0 | 1.6 | 53.3 | 1.1 |
| *K. lactis (I)* | 0[b] | 312 | 0[c] | 55 | 256 | 16 | 27 | 65 | 43.4 | 2.6 | 52.1 | 1.9 |
| *K. thermotolerans (E)* | 16[a] | 304 | 2 | 44 | 204 | 17 | 2[d] | 56 | 47.7 | 0.9 | 51.0 | 0.4 |
| *K. thermotolerans (I)* | 42[a] | 440 | 10 | 72 | 298 | 23 | 30[d] | 78 | 48.3 | 2.1 | 47.6 | 2.0 |
| *C. glabrata (E)* | 11[a] | 294 | 1[c] | 45 | 207 | 21 | 16[d,e] | 73 | 46.7 | 0.6 | 52.0 | 0.7 |
| *C. glabrata (I)* | 28[a] | 415 | 1[c] | 60 | 318 | 25 | 85[d,e] | 78 | 48.6 | 0.9 | 49.8 | 0.7 |
| *S. cerevisiae(E)* | 33[a] | 333 | 7 | 49 | 239 | 31 | 60[d,e] | 73 | 48.6 | 2.3 | 47.1 | 2.0 |
| *S. castellii (E)* | 19[a] | 274 | 0[c] | 40 | 203 | 7 | 101[d,e] | 56 | 47.5 | 1.6 | 49.9 | 1.0 |
| *S. servazzii (E)* | 22[a] | 300 | 0[c] | 46 | 218 | 11 | 95[d,e] | 70 | 37.2 | 0.9 | 59.5 | 2.4 |

Table 4: Codon usage in some Metazoan lineages.
[a]AUA is Met in these species; [b] AGR remains Arg; [c] AGR is Gly; [d] AGR is Stop;
[e] AGG is avoided, probably because there is an unmodified G in the wobble position of tRNA-Ser(GCU);
[f] AAA is reassigned to Asn.

| | Ile and Met codons | | | | Ser and Arg codons | | | | Asn and Lys codons | | | | Frequency at FFD sites | | | |
|---|---|---|---|---|---|---|---|---|---|---|---|---|---|---|---|---|
| | AUU | AUC | AUA | AUG | AGU | AGC | AGA | AGG | AAU | AAC | AAA | AAG | %U | %C | %A | %G |
| *Axinella corrugata* | 222 | 45 | 205 | 130 | 74 | 23 | 35[b] | 17 | 107 | 14 | 86 | 28 | 35.9 | 11.7 | 30.6 | 21.8 |
| *Geodia neptuni* | 207 | 18 | 218 | 118 | 79 | 15 | 55[b] | 15 | 104 | 18 | 89 | 24 | 40.7 | 7.1 | 35.7 | 16.4 |
| *Metridium senile* | 190 | 59 | 110 | 138 | 58 | 21 | 43[b] | 12 | 84 | 31 | 79 | 22 | 42.5 | 17.5 | 25.0 | 15.0 |
| *Acropora tenuis* | 182 | 44 | 115 | 113 | 68 | 11 | 34[b] | 16 | 90 | 15 | 72 | 31 | 43.5 | 11.5 | 17.6 | 27.4 |
| *Limulus polyphemus* | 241 | 107 | 171[a] | 43 | 22 | 8 | 66 | 12 | 103 | 44 | 66 | 19 | 35.7 | 17.1 | 38.8 | 8.4 |
| *Daphnia pulex* | 187 | 93 | 90[a] | 52 | 49 | 20 | 68 | 1[e] | 75 | 43 | 48 | 41 | 36.2 | 20.2 | 24.9 | 18.7 |
| *Drosophila melanogaster* | 355 | 16 | 216[a] | 13 | 30 | 0 | 74 | 0[e] | 193 | 10 | 81 | 5 | 50.5 | 2.4 | 43.6 | 3.6 |
| *Caenorhabditis elegans* | 257 | 23 | 134[a] | 44 | 61 | 6 | 126 | 39 | 139 | 12 | 95 | 14 | 50.2 | 4.1 | 37.5 | 8.2 |
| *Trichinella spriralis* | 136 | 114 | 193[a] | 96 | 46 | 20 | 71 | 34 | 68 | 80 | 65 | 34 | 29.6 | 15.2 | 42.2 | 13.0 |
| *Katharina tunicata* | 226 | 49 | 125[a] | 45 | 46 | 36 | 88 | 35 | 134 | 47 | 79 | 30 | 43.1 | 12.3 | 32.0 | 12.5 |
| *Lumbricus terrestris* | 194 | 105 | 194[a] | 64 | 20 | 14 | 63 | 13 | 70 | 66 | 70 | 22 | 28.2 | 24.3 | 36.9 | 10.5 |
| *Terebratulina retusa* | 137 | 147 | 152[a] | 34 | 6 | 23 | 78 | 18 | 46 | 67 | 80 | 19 | 22.6 | 35.3 | 34.4 | 7.7 |
| *Fasciola hepatica* | 119 | 8 | 39 | 96 | 92 | 8 | 7 | 41 | 50 | 2 | 18[f] | 44 | 68.2 | 5.4 | 6.9 | 19.4 |
| *Schistosoma mansoni* | 134 | 15 | 155 | 123 | 102 | 10 | 62 | 50 | 54 | 4 | 39[f] | 60 | 52.8 | 4.3 | 23.3 | 19.5 |
| *Taenia crassiceps* | 174 | 2 | 126 | 101 | 104 | 1 | 43 | 29 | 89 | 5 | 59[f] | 49 | 64.3 | 1.4 | 19.9 | 14.5 |
| *Paracentrotus lividus* | 146 | 53 | 165 | 102 | 11 | 14 | 66 | 15 | 44 | 50 | 88[f] | 54 | 25.0 | 22.9 | 42.6 | 9.6 |
| *Asterina pectinifera* | 129 | 66 | 178 | 78 | 22 | 23 | 54 | 19 | 40 | 58 | 110[f] | 48 | 28.8 | 25.1 | 35.4 | 10.6 |
| *Balanoglossus carnosus* | 97 | 148 | 69 | 77 | 18 | 35 | 19 | 0[e] | 20 | 98 | 0 | 45 | 22.4 | 37.8 | 29.1 | 10.6 |
| *Saccoglossus kowalewskii* | 185 | 72 | 96 | 70 | 34 | 22 | 4 | 0[e] | 60 | 68 | 47 | 12 | 40.7 | 23.0 | 31.2 | 5.1 |
| *Halocynthia roretzi* | 170 | 12 | 112[a] | 60 | 125 | 4 | 68[c] | 88 | 78 | 5 | 29 | 42 | 53.0 | 3.6 | 26.1 | 17.2 |
| *Cionia intestinalis* | 297 | 19 | 237[a] | 34 | 73 | 7 | 221[c] | 37 | 143 | 16 | 103 | 22 | 56.5 | 4.8 | 33.3 | 5.4 |
| *Branchiostoma lanceolatum* | 186 | 47 | 139[a] | 63 | 78 | 30 | 12 | 0[e] | 86 | 26 | 40 | 34 | 37.7 | 9.1 | 32.8 | 20.5 |
| *Branchiostoma floridae* | 187 | 46 | 138[a] | 63 | 73 | 33 | 12 | 0[e] | 87 | 24 | 40 | 33 | 38.5 | 8.5 | 33.0 | 20.0 |
| *Epigonichthys maldivensis* | 159 | 70 | 122[a] | 59 | 49 | 35 | 31 | 2[e] | 70 | 40 | 37 | 35 | 34.7 | 13.6 | 29.4 | 22.3 |
| *Myxine glutinosa* | 210 | 154 | 182[a] | 44 | 23 | 28 | 2[d] | 1 | 69 | 77 | 102 | 15 | 29.2 | 27.4 | 34.9 | 8.5 |
| *Homo sapiens* | 124 | 196 | 167[a] | 40 | 14 | 39 | 1[d] | 1 | 32 | 132 | 85 | 10 | 14.5 | 40.4 | 38.7 | 6.4 |

Table 5: Codon usage in some green algae. Intronic and other non-standard ORFs are excluded. Unusual selective effects influence *S. obliquus* and *C. reinhardtii*.

| | Leu codons | | Ile | Ser codons | | Thr codons | | Glu codons | | Arg codons | | FFD sites | |
|---|---|---|---|---|---|---|---|---|---|---|---|---|---|
| | UUA | UUG | AUA | UCA | UCG | ACA | ACG | GAA | GAG | AGA | AGG | %A | %G |
| *Pedinomonas minor* | 411 | 84 | 115 | 53 | 8 | 27 | 0 | 58 | 11 | 35 | 2 | 21.3 | 1.6 |
| *Scenedesmus obliquus* | 0 | 384 | 0 | 13 | 0 | 111 | 5 | 73 | 14 | 0 | 93 | 35.7 | 6.6 |
| *Chlamydomonas eugametos* | 353 | 38 | 176 | 80 | 3 | 98 | 6 | 53 | 4 | 23 | 0 | 48.5 | 11.7 |
| *Chlamydomonas reinhardtii* | 0 | 251 | 0 | 0 | 0 | 0 | 0 | 0 | 55 | 0 | 0 | 26.1 | 3.7 |

**Table 6**

| Codon reassignment | No. of times | Can this be explained by GC→AU mutation pressure? | Change in No. of tRNAs | Is mispairing important? | Reassignment Mechanism |
|---|---|---|---|---|---|
| UAG: Stop → Leu | 2 | Yes. G → A at 3rd position. | +1 | No | CD |
| UAG: Stop → Ala | 1 | Yes. G → A at 3rd position | +1 | No | CD |
| UGA: Stop → Trp | 12 | Yes. G → A at 2nd position. | 0 | Possibly. CA at 3rd position. | CD |
| CGU/CGC/CGA/CGC: Arg → Unassigned | 5 | Yes. C → A at 1st position. | -1 | No | CD |
| CUU/CUC/CUA/CUG: Leu → Thr | 1 | Yes. C → U at 1st position. | 0 | No | CD |
| CUU/CUC/CUA/CUG: Thr → Unassigned | 1 | No | -1 | No | CD |
| AUA: Ile → Met or Unassigned | 3 / 5[a] | No | -1 | Yes. GA at 3rd position. | UC |
| AAA: Lys → Asn | 2 | No | 0 | Yes. GA at 3rd position. | AI |
| AAA: Lys → Unassigned | 1 | No | 0 | Possibly. GA at 3rd position. | UC or AI |
| AGA/AGG: Arg → Ser | 1 | No | -1 | Yes. GA at 3rd position. | UC |
| AGA/AGG: Ser → Stop | 1 | No | 0 | No | AI[b] |
| AGA/AGG: Ser → Gly | 1 | No | +1 | No | AI[b] |
| UUA: Leu → Stop | 1 | No | 0 | No | UC or AI |
| UCA: Ser → Stop | 1 | No | 0 | No | UC or AI |

[a] The loss of the tRNA-Ile has occurred 3 times (once in metazoa, once in yeasts, and once in *S. octosporus*) but the reassignment of the codon to Met has occurred at least 5 times (3 times in metazoa and twice in yeasts).
[b] If AGR was fully reassigned to Ser prior to these changes, then they are probably AI. However, they may also be considered as a final response to the loss of the original tRNA-Arg, in which case we might count them as UC.

**Figure Captions**

1. Phylogeny of fungi and related species derived from mitochondrial proteins.

2. Phylogeny of plants and algae derived from mitochondrial proteins.

3. Phylogeny of alveolates, stramenopiles and haptophytes according to published sources.

4. Phylogeny of metazoa according to published sources.

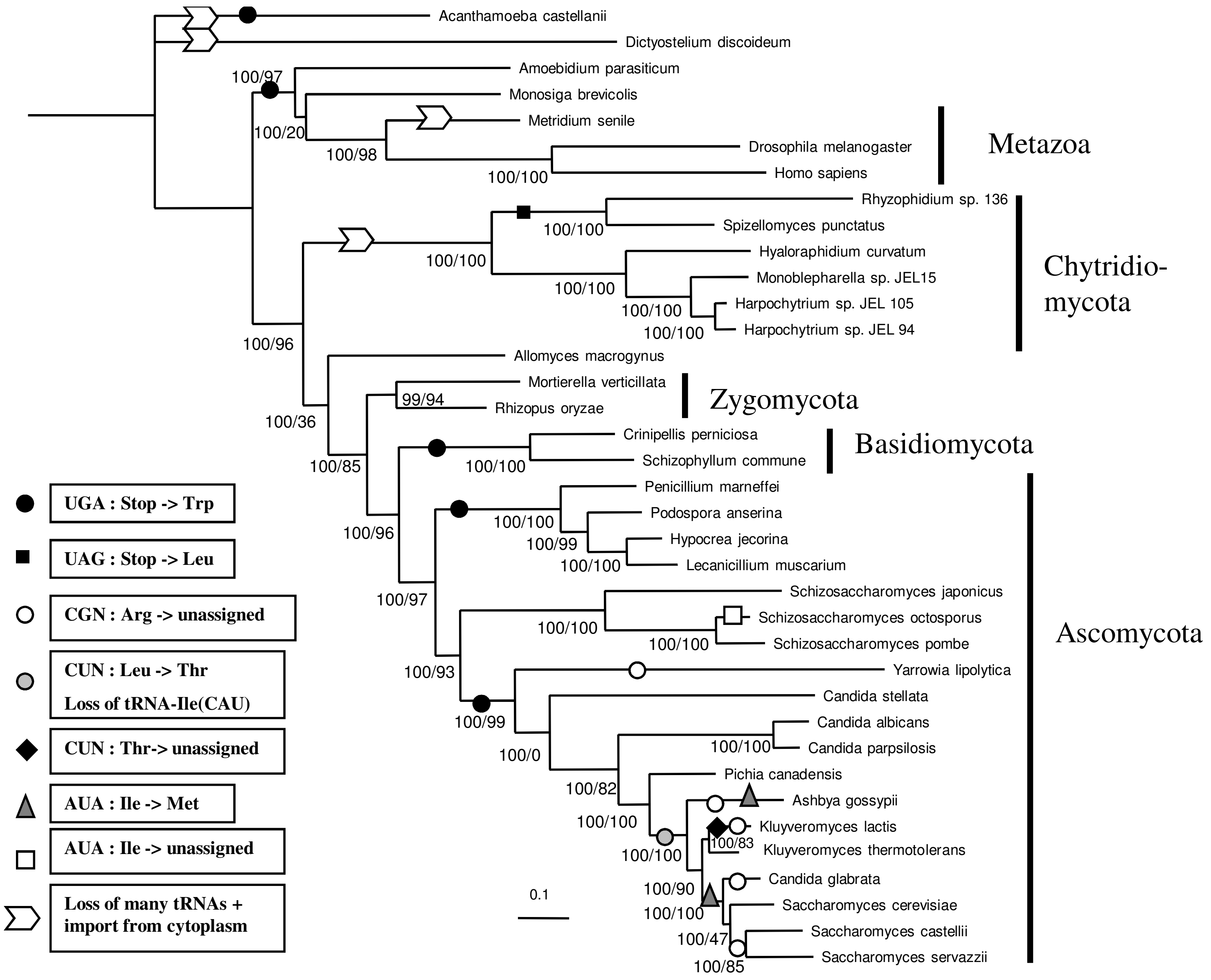

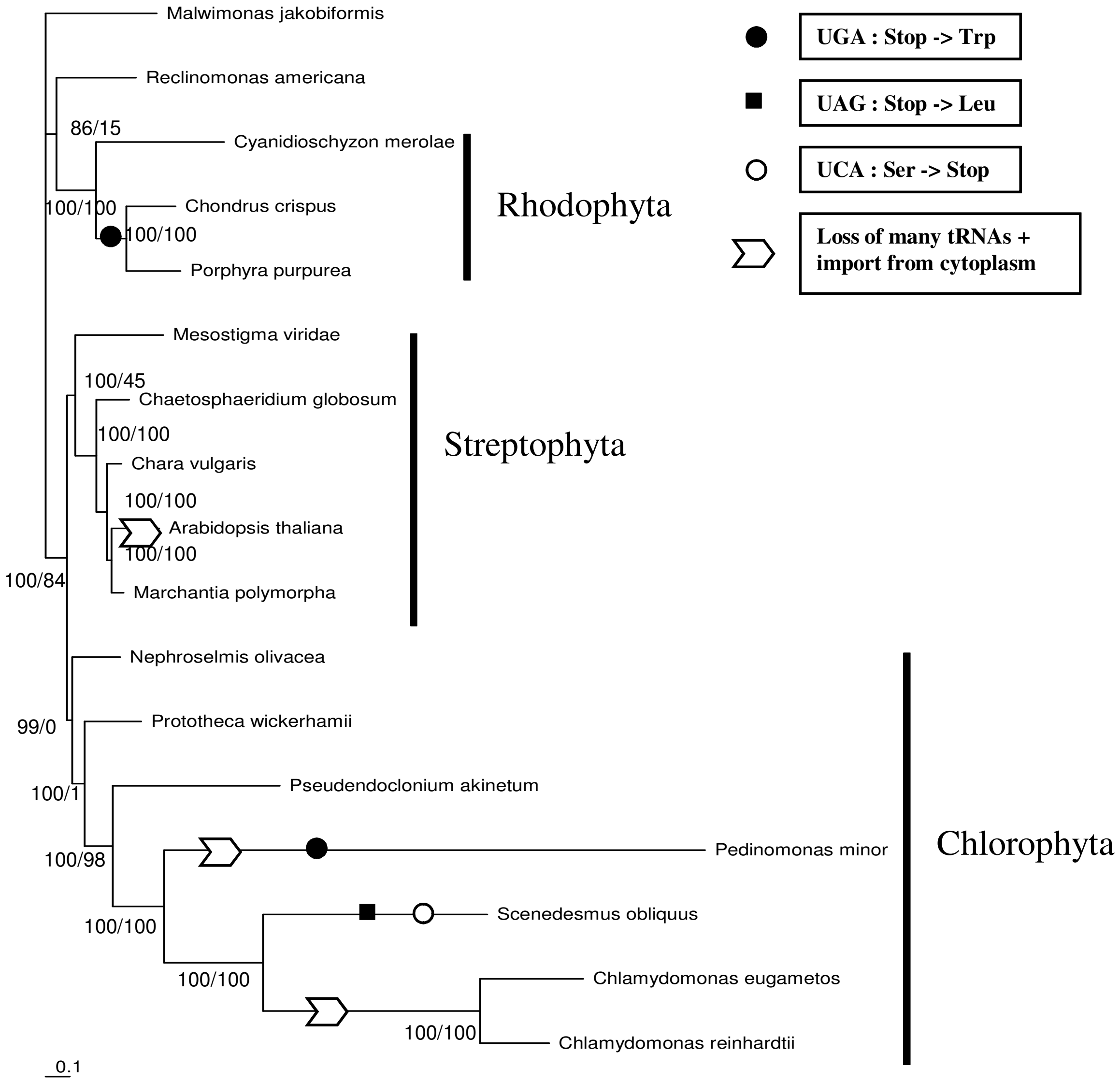
Malwimonas jakobiformis
Reclinomonas americana
86/15
Cyanidioschyzon merolae
100/100
Chondrus crispus
100/100
Porphyra purpurea
Rhodophyta
Mesostigma viridae
100/45
Chaetosphaeridium globosum
100/100
Chara vulgaris
100/100
Arabidopsis thaliana
100/100
Marchantia polymorpha
100/84
Streptophyta
Nephroselmis olivacea
99/0
Prototheca wickerhamii
100/1
Pseudendoclonium akinetum
100/98
Pedinomonas minor
100/100
Scenedesmus obliquus
100/100
Chlamydomonas eugametos
100/100
Chlamydomonas reinhardtii
Chlorophyta
0.1
UGA : Stop -> Trp
UAG : Stop -> Leu
UCA : Ser -> Stop
Loss of many tRNAs + import from cytoplasm

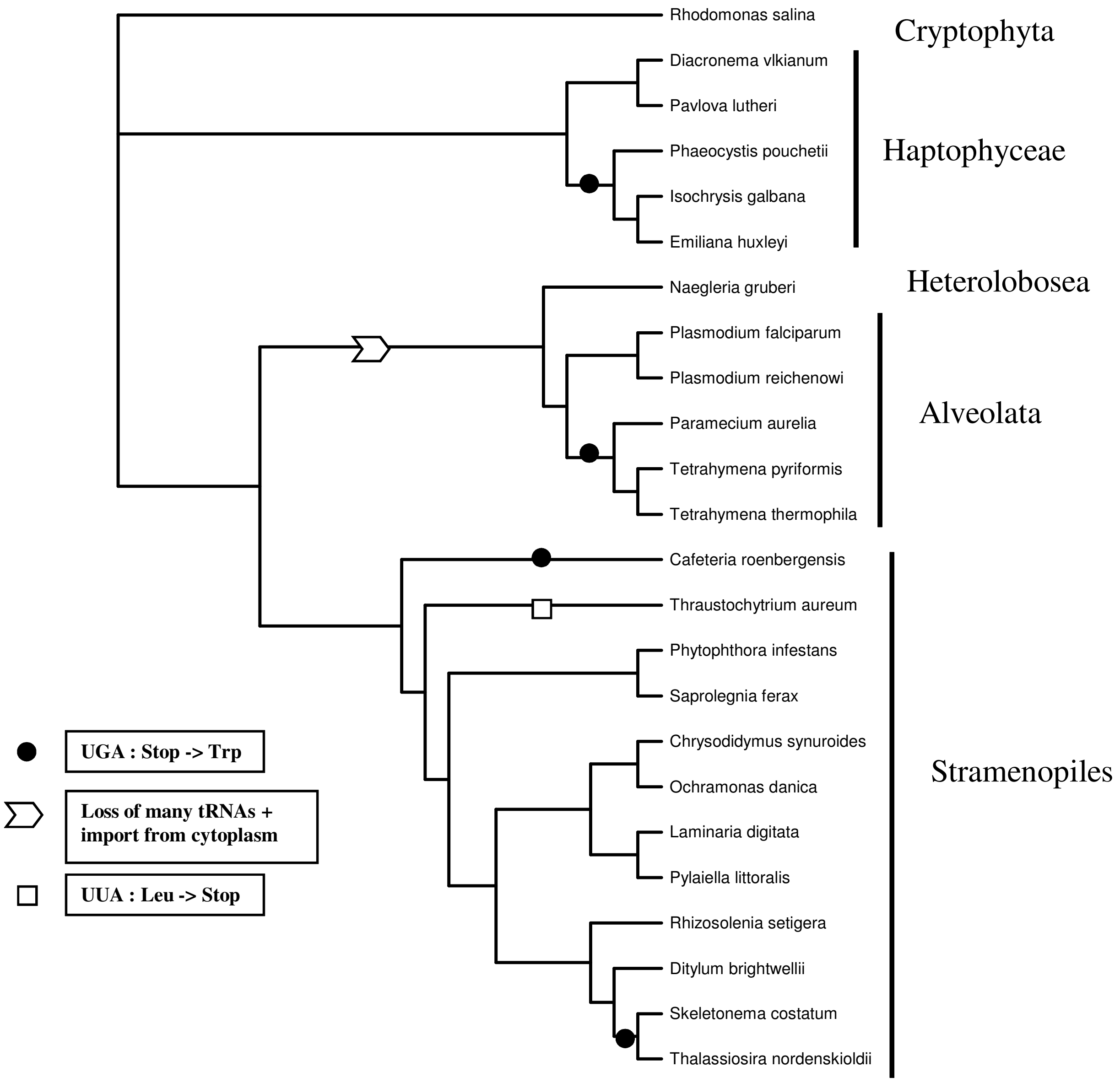
Rhodomonas salina
Cryptophyta
Diacronema vlkianum
Pavlova lutheri
Phaeocystis pouchetii
Haptophyceae
Isochrysis galbana
Emiliana huxleyi
Naegleria gruberi
Heterolobosea
Plasmodium falciparum
Plasmodium reichenowi
Paramecium aurelia
Alveolata
Tetrahymena pyriformis
Tetrahymena thermophila
Cafeteria roenbergensis
Thraustochytrium aureum
Phytophthora infestans
Saprolegnia ferax
Chrysodidymus synuroides
Stramenopiles
Ochramonas danica
Laminaria digitata
Pylaiella littoralis
Rhizosolenia setigera
Ditylum brightwellii
Skeletonema costatum
Thalassiosira nordenskioldii
UGA : Stop -> Trp
Loss of many tRNAs + import from cytoplasm
UUA : Leu -> Stop

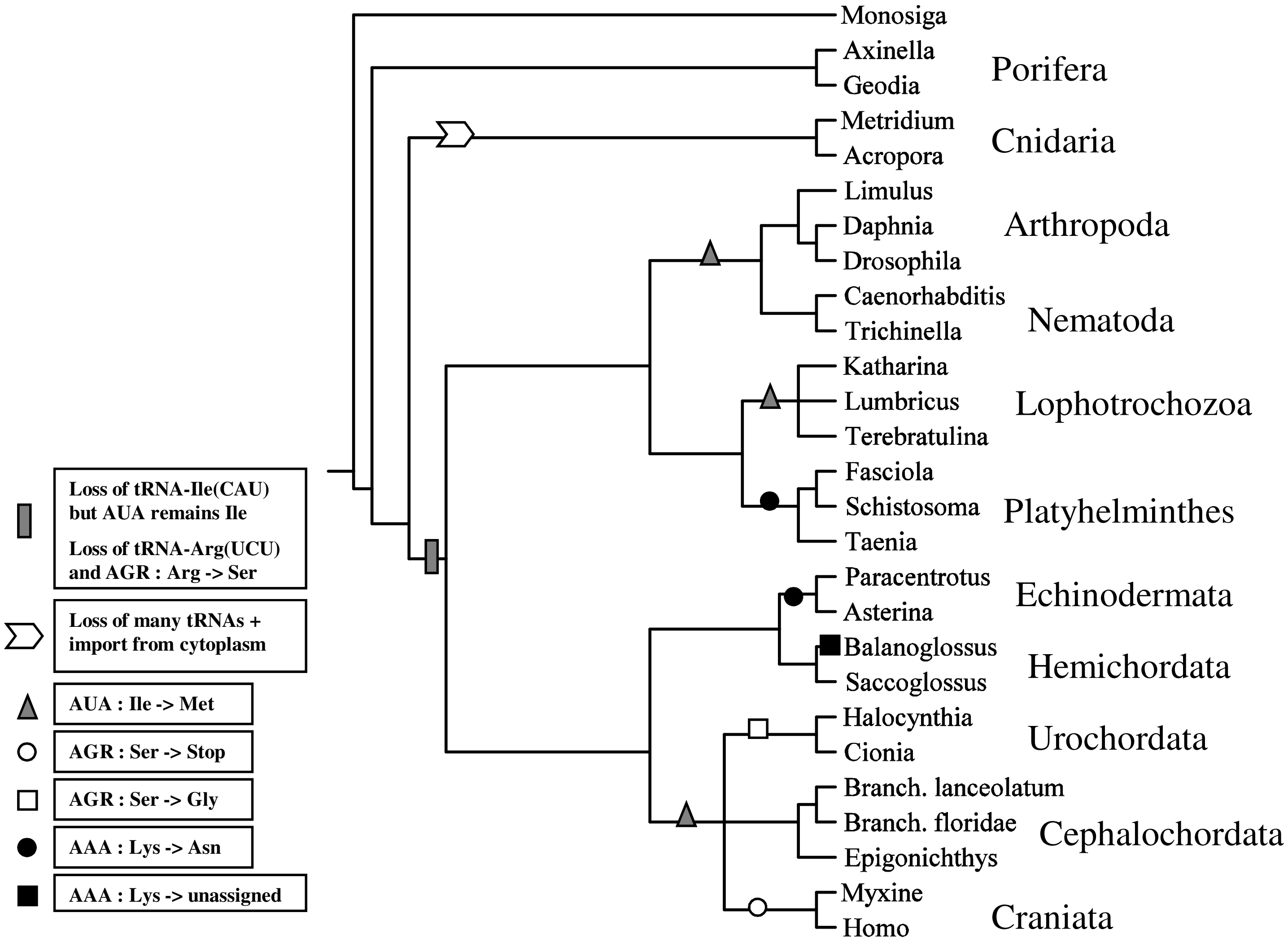
Monosiga
Axinella
Geodia
Porifera
Metridium
Acropora
Cnidaria
Limulus
Daphnia
Drosophila
Arthropoda
Caenorhabditis
Trichinella
Nematoda
Katharina
Lumbricus
Terebratulina
Lophotrochozoa
Fasciola
Schistosoma
Taenia
Platyhelminthes
Paracentrotus
Asterina
Echinodermata
Balanoglossus
Saccoglossus
Hemichordata
Halocynthia
Cionia
Urochordata
Branch. lanceolatum
Branch. floridae
Epigonichthys
Cephalochordata
Myxine
Homo
Craniata
Loss of tRNA-Ile(CAU) but AUA remains Ile
Loss of tRNA-Arg(UCU) and AGR : Arg -> Ser
Loss of many tRNAs + import from cytoplasm
AUA : Ile -> Met
AGR : Ser -> Stop
AGR : Ser -> Gly
AAA : Lys -> Asn
AAA : Lys -> unassigned

# The Mechanisms of Codon Reassignments in Mitochondrial Genetic Codes

Supratim Sengupta, Xiaoguang Yang, and Paul G. Higgs

## Supplementary Information

### Methods

The list of species used in our study in each of the principal eukaryotic taxa is given in Table S1. In taxa with few complete mitochondrial genomes, all the available species were included. In taxa with many complete genomes (particularly metazoa), representative species were used. Table S1 also gives information on the set of amino acids for which tRNAs are present in each genome, and the reference number of the genetic code used by each species following the NCBI genetic code numbering system. This system is summarized in Table S2 for reference.

For the plants/algae set we used concatenated protein alignments of cox1,2,3,cob,atp6,9,nad1,3,4,4l,5, having a total length of 3385 amino acids. The genes for atp8, nad2 and nad6 were found to be substantially more divergent than the rest, and were excluded to avoid introducing noise into the data. For the fungi, most of the nad genes are missing. Therefore, we used an alignment of cox1,2,3,cob,atp6,9 with a total length of 1701 amino acids. In a few cases when a specific gene was missing in any species (e.g. *Candida albicans* does not have the cob gene), the entry for that gene was filled by gaps. Multiple sequence alignments were carried out using T-COFFEE (Notredame *et al.* 2000) and the alignment was edited to remove columns having more than 20% gaps. Phylogenetic analysis was performed using the MCMC program available in the PHASE package (Jow *et al.* 2003). The mtREV24 amino acid substitution matrix with a gamma distribution for site variation with four categories was used. In order to compare the Bayesian posterior probabilities from MCMC with bootstrap values, we also generated 100 bootstrap replicate data sets. Maximum likelihood pairwise distance matrices were calculated for each replicate using the same evolutionary model, and trees were constructed using the neighbour joining (NJ) method using the PHYLIP package (Felsenstein 1989). Bootstrap percentages were obtained from this set of trees.

We also carried out independent phylogenetic analysis with a combined alignment of large and small subunit rRNA genes to ascertain if the trees derived from the proteins and rRNAs are consistent with one another. The rRNA sequences were aligned initially using ClustalX (Thompson *et al.* 1997). Variable regions of the molecules were excluded from the analysis by deleting sites with more than 10% gaps in the alignment. The lengths of the alignments were 2063 and 3829 for the fungi and plants/algae. We analyzed the rRNA data sets using a method that accounts for conserved RNA structure. A consensus secondary structure was added to the alignments. The general reversible model was used for the unpaired sites, and a model specifically treating compensatory substitutions was used for the paired sites (model 7A in Savill *et al.* 2001). Parameters for both models are optimized simultaneously by the MCMC program (see Hudelot *et al.* 2003 and the documentation to the PHASE package available at http://www.bioinf.man.ac.uk/resources/phase/). For both models, variation of rates across sites was accounted four using 4 gamma-distributed categories.

## Phylogenetic discussion

Although there is general agreement on the definitions of the major eukaryotic taxa, the relationships between groups in the very earliest parts of the eukaryotic tree have proved difficult to resolve (Lang *et al.* 1999; Baldauf *et al.* 2000; Van der Peer *et al.* 2000; Baldauf 2003; Philippe *et al.* 2004). Our aim here was not to resolve these early branches, but to obtain trees that would allow for accurate positioning of the codon reassignments.

The grouping of fungi and metazoa, proposed by Baldauf & Palmer (1993), is supported by all the studies cited in the previous paragraph. *Amoebidium parasiticum* and *Monosiga brevicollis* are thought to be the closest unicellular relatives of the metazoa, and are therefore also included in this group. *Dictyostelium discoideum* and *Acanthamoeba castellanii* are thought to be more distant relatives of this group that diverge before the split of fungi and metazoa (Baldauf *et al.* 2000; Van der Peer *et al.* 2000; Lang *et al.* 2002). These are therefore included in this subset as outgroups.

In Figure 1 we focus on the phylogeny within the fungi and include only a few representative metazoa. The tree from the mitochondrial proteins is very well resolved. Almost all nodes have 100% posterior probability (PP), and most also have high bootstrap percentage (BP). The branching order of the species within Ascomycota agrees with Bullerwell *et al.* (2003a) and Kurtzman and Robnett (2003), and that within Chytridiomycota agrees with Bullerwell *et al.* (2003b). An exception is *A. macrogynus*. In the NJ analysis, the most common position of this species is with the Chytridiomycota, but it appears as basal to the Zygomycota/Basidiomycota/Ascomycota according the MCMC analysis in Figure 1. When the Zygomycota are excluded, *A. macrogynus* moves to the base of the Chytridiomycota in the MCMC tree. The uncertainty in the position of this species does not affect any of our conclusions regarding changes in the genetic code.

The mitochondrial rRNA tree topology for the fungi is identical to the protein tree in Figure 1, with the exception of the position of *Candida stellata*. In the rRNA tree *C. stellata* pairs with *Yarrowia* (PP = 98%, BP = 72%). In the protein tree, *C. stellata* branches after *Yarrowia* (PP = 100%), although the BP is 0% for this arrangement because in the protein NJ consensus tree, it branches before *Yarrowia*. The positioning of *C. stellata* is therefore not clear. Nevertheless, whatever its position, *C. stellata* is separate from *C. albicans* + *C. parapsilosis*, and these are all separate from *C. glabrata*. Thus, it is clear that *Candida* is paraphyletic. Kurzman & Robnett (2003) and Diezmann *et al.* (2004) have considered the phylogeny of these groups using nuclear genes, and find that *Candida*, *Kluyveromyces, Saccharomyces* and *Pichia* are all paraphyletic.

The second data set contains Rhodophyta (red algae), Chlorophyta (green algae) and Streptophyta (plants). *Malwimonas jakobiformis* and *Reclinomonas americana* are added as outgroups. These are found to be related to the plants and algae by Lang *et al.* (2002) and Forget *et al.* (2002). The protein sequence phylogeny for this group (Figure 2) is again quite well resolved. The Rhodophyta, Chlorophyta and Streptophyta are all monophyletic, with Rhodophyta basal to the other two, which is consistent with results of Lang *et al.* (2002). However, the grouping of *Reclinomonas* with the rhodophytes, instead of being an outgroup to rhodophytes + chlorophytes + streptophytes is most likely an artefact. The positioning of *Nephroselmis* and *Prototheca* is also somewhat questionable because these species appear in different positions in the NJ tree and hence they have low BP in Figure 2. The rRNA phylogeny is the same as Figure 2 with the following exceptions. In this case *Reclinomonas* and *Malawimonas* are indeed outgroups as expected, and there is strong support for rhodophytes + chlorophytes + streptophytes (100% PP). However, in the rRNA tree, *Nephroselmis* shifts to being basal to Streptophyta instead of being basal to the other Chlorophyta. Turmel *et al.* (1999)

find that the position of *Nephroselmis* shifts according to the method and sequences used, but conclude that it is the basal species within Chlorophyta. Figure 2 is also equivalent to the tree obtained by Pombert *et al.* (2004). We therefore conclude that, for the study in the remainder of this paper, the protein tree in Figure 2 is the best estimate of the phylogeny of this group, with the interpretation that *Reclinomonas* should be an outgroup. The position of *Reclinomonas*, and the uncertainty regarding *Nephroselmis* do not affect the conclusions regarding codon reassignments that we make later in the paper.

The third data set contains species that are clearly not associated with either the fungi/metazoa or plants/algae. Basal branches in this tree have been left as multifurcating. This set may not be monophyletic with respect to the groups in Figures 1 and 2, *i.e.* it is possible that the fungi/metazoa or plants/algae may lie within the tree of Figure 3. If this is the case, it does not affect the conclusions regarding the genetic code. In order to position some of the genetic code changes in this set it is necessary to include species for which complete mitochondrial genomes are not available. We therefore cannot not show a tree derived only from mitochondrial genes. Stramenopiles and alveolates are two groups that are thought to be monophyletic according to Wolters (1991) and Saunders *et al.* (1995). The combination of these two groups is sometimes called Chromalveolata, and is supported by Van der Peer *et al.* (2000), Baldauf *et al.* (2000), and Philippe *et al.* (2004). For the arrangement of the species within the Stramenopiles, we used the results of a much larger scale study with the nuclear small subunit rRNA gene (Van der Peer *et al.* 1996). For the species in which mitochondrial genomes were available, we carried out protein and rRNA phylogenies as with the previous two sets. This was not very informative due to the small number of species available. However, one useful point that emerged was that *N. gruberi* appeared as sister group to the Alveolates with 100% PP from both the protein and rRNA analysis. We have therefore placed it in that position on Figure 3. We are not confident of this result due to the poor species sampling and the possibility of long-branch artefacts involving the Alveolates. Repositioning this species might lead to an increase in the inferred number of times that tRNA import has evolved, but it would not affect the positioning of the codon reassignments. The Haptophyceae are a separate group from Alveolates and Stramenopiles. For the species in this group we followed Hayashi-Ishimaru et.al. (1997). The complete genome of *E. huxleyi* has become available since that study, and has been added in Figure 3. We confirmed its relationship to *I. galbana* using the cox1 protein sequence only.

Figure 4 shows the metazoa. This is a well studied group. We have followed Mallatt & Winchell (2002) and Halanych (2004) for the relationship between the animal phyla.

Table S1 - List of species with complete mitochondrial genomes used in this study

| Species | Taxon | Code No. | Amino acids for which tRNAs are present |
|---|---|---|---|
| **Figure 1** | | | |
| Acanthamoeba castellanii | Acanthamoebidae | 4 | A, D, E, F, H, I, K, L(CUN), L(UUR), M, P, Q, W, Y |
| Dictyostelium discoideum | Mycetozoa | 1 | All but D, G, R(CGN), S(UCN), S(AGY), T, V |
| Amoebidium parasiticum | Ichthyosporia | 4 | Full set |
| Monosiga brevicollis | Choanoflagellida | 4 | Full set |
| Rhizophydium sp. 136 | Chytridiomycota | 16 | K, L(UAG), M, P, Q, W, Y |
| Spizellomyces punctatus | Chytridiomycota | 16 | D, K, L(UAG), M, P, Q, W, Y |
| Hyaloraphidium curvatum | Chytridiomycota | 1* | D, E, M, P, Q, W, Y |
| Monoblepharella sp. JEL15 | Chytridiomycota | 1 | D, E, G, K, M, P, Q, W, Y |
| Harpochytrium sp. JEL105 | Chytridiomycota | 1* | D, E, K, M, P, Q, W, Y |
| Harpochytrium sp. JEL94 | Chytridiomycota | 1* | D, E, K, M, P, Q, W, Y |
| Allomyces macrogynus | Chytridiomycota | 1* | Full set |
| Mortierella verticillata | Zygomycota | 1* | Full set |
| Rhizopus oryzae | Zygomycota | 1 | Full set |
| Crinipellis perniciosa | Basidiomycota | 4 | Full set |
| Schizophyllum commune | Basidiomycota | 4 | Full set |
| Penicillium marneffei | Ascomycota | 4 | Full set |
| Podospora anserina | Ascomycota | 4 | Full set |
| Hypocrea jecorina | Ascomycota | 4 | Full set |
| Lecanicillium muscarium | Ascomycota | 4 | All but A, C |
| Schizosaccharomyces japonicus | Ascomycota | 4 | Full set |
| Schizosaccharomyces octosporus | Ascomycota | 4 | Full set |
| Schizosaccharomyces pombe | Ascomycota | 4 | Full set |
| Yarrowia lipolytica | Ascomycota | 4 | All but R(CGN), V |
| Candida stellata | Ascomycota | 4 | Full set |
| Candida albicans | Ascomycota | 4 | Full set |
| Candida parapsilosis | Ascomycota | 4 | Full set |
| Pichia canadensis | Ascomycota | 4 | Full set |
| Ashbya gossypii | Ascomycota | 3 | All but Q, R(CGN) |
| Kluyveromyces lactis | Ascomycota | 3 | All but L(CUN), R(CGN) |
| Kluyveromyces thermotolerans | Ascomycota | 3 | All but L(CUN) |
| Candida glabrata | Ascomycota | 3 | All but L(CUN), R(CGN) |
| Saccharomyces cerevisiae | Ascomycota | 3 | All but L(CUN) |
| Saccharomyces castellii | Ascomycota | 3 | All but L(CUN), R(CGN) |
| Saccharomyces servazzii | Ascomycota | 3 | All but L(CUN), R(CGN) |
| **Figure 2** | | | |
| Malwimonas jakobiformis | Malawimonadidae | 1 | Full set |
| Reclinomonas americana | Jakobidae | 1 | All but T |
| Cyanidioschyzon merolae | Rhodophyta | 1 | All but T |
| Chondrus crispus | Rhodophyta | 4 | All but S(AGY), T |
| Porphyra purpurea | Rhodophyta | 4 | All but T |
| Mesostigma viride | Streptophyta | 1 | All but T |
| Chaetosphaeridium globosum | Streptophyta | 1 | Full set |
| Chara vulgaris | Streptophyta | 1 | Full set |
| Marchantia polymorpha | Streptophyta | 1 | All but I |
| Arabidopsis thaliana | Streptophyta | 1 | C, D, F, G, H, K, M, N, P, Q, S(UCN), S(AGY), W, Y |
| Nephroselmis olivacea | Chlorophyta | 1 | Full set |
| Prototheca wickerhamii | Chlorophyta | 1 | Full set |
| Pseudendoclonium akinetum | Chlorophyta | 1 | Full set |
| Pedinomonas minor | Chlorophyta | 4 | C, E, F, H, L(UUR), Q, W, Y |
| Scenedesmus obliquus | Chlorophyta | 22 | All but T |
| Chlamydomonas eugametos | Chlorophyta | 1 | M, Q, W |
| Chlamydomonas reinhardtii | Chlorophyta | 1 | M, Q, W |
| **Figure 3** | | | |
| Rhodomonas salina | Cryptophyta | 1 | Full set |
| Emiliania huxleyi | Haptophyceae | 4 | Full set |
| Naegleria gruberi | Heterolobosea | 1 | All but A, C, G, R(CGN), T, V |
| Plasmodium falciparum | Alveolata | 1* | None |
| Plasmodium reichenowi | Alveolata | 1* | None |
| Paramecium aurelia | Alveolata | 4 | F, M, W, Y |
| Tetrahymena pyriformis | Alveolata | 4 | F, H, L(UUR), M, Q, W, Y |
| Tetrahymena thermophila | Alveolata | 4 | F, H, L(UUR), M, Q, W, Y |
| Cafeteria roenbergensis | Stramenopiles | 4 | All but L(CUN), R(CGN), T |
| Thraustochytrium aureum | Stramenopiles | 23 | All but R(CGN), R(AGR), T |
| Phytophthora infestans | Stramenopiles | 1 | All but T |
| Saprolegnia ferax | Stramenopiles | 1 | All but T |
| Chrysodidymus synuroideus | Stramenopiles | 1 | All but R(CGN), T |
| Ochromonas danica | Stramenopiles | 1 | All but R(CGN), T |
| Laminaria digitata | Stramenopiles | 1 | All but R(CGN), T |
| Pylaiella littoralis | Stramenopiles | 1 | All but R(CGN), T |
| **Figure 4** | | | |
| Axinella corrugata | Porifera | 4 | Full set |
| Geodia neptuni | Porifera | 4 | Full set |
| Metridium senile | Cnidaria | 4 | M, W |
| Acropora tenuis | Cnidaria | 4 | M, W |
| Limulus polyphemus | Arthropoda | 5 | Full set |
| Daphnia pulex | Arthropoda | 5 | Full set |
| Drosophila melanogaster | Arthropoda | 5 | Full set |
| Caenorhabditis elegans | Nematoda | 5 | Full set |
| Trichinella spiralis | Nematoda | 5 | Full set |
| Katharina tunicata | Mollusca | 5 | Full set |
| Lumbricus terrestris | Annelida | 5 | Full set |
| Terebratulina retusa | Brachiopoda | 5 | Full set |
| Fasciola hepatica | Platyhelminthes | 9 | Full set |
| Schistosoma mansoni | Platyhelminthes | 9 | Full set |
| Taenia crassiseps | Platyhelminthes | 9 | Full set |
| Paracentrotus lividus | Echinodermata | 9 | Full set |
| Asterina pectinifera | Echinodermata | 9 | Full set |
| Balanoglossus carnosus | Hemichordata | 9 | Full set |
| Saccoglossus kowalewskii | Hemichordata | 9a | Full set |
| Halocynthia roretzi | Urochordata | 13 | Full set |
| Cionia intestinalis | Urochordata | 13 | Full set |
| Branchiostoma lanceolatum | Cephalochordata | 5 | Full set |
| Branchiostoma floridae | Cephalochordata | 5 | Full set |
| Epigonichthys maldivensis | Cephalochordata | 5 | Full set |
| Myxine glutinosa | Craniata | 2 | Full set |
| Homo sapiens | Vertebrata | 2 | Full set |
| Didelphis virginiana | Vertebrata | 2 | All but K |

Table S2 - Summary of Genetic Code Numbering system (following NCBI)

| Code No. | Differences from Canonical Code |
|---|---|
| 1 | Canonical Code. (Species marked 1* use the canonical code but are erroneously labelled in NCBI.) |
| 4 | UGA = Trp. |
| 5 | UGA = Trp; AUA = Met; AGR = Ser. |
| 2 | UGA = Trp; AUA = Met; AGR = Stop. |
| 11 | UGA = Trp; AUA = Met; AGR = Gly. |
| 21 | UGA = Trp; AUA = Met; AGR = Ser; AAA = Asn. |
| 3 | UGA = Trp; AUA = Met or unassigned; CGN = rare or unassigned; CUN = Thr. |
| 16 | UAG = Leu. |
| 22 | UAG = Leu; UCA = Stop. |
| 23 | UUA = Stop. |
| 9 | UGA = Trp; AGR = Ser; AAA = Asn or unassigned |
| 9a | UGA = Trp; AGR = Ser. (This is similar to 9 but does not have a number in the NCBI system) |

----------------------------------------------------------------------------------------------------------

Table S3: Annotation of tRNA's with CAU anticodon in the Fungi group together with their position on the mitochondrial genome. *Podospora anserina* has two identical tRNA-Met(CAU) elongator genes labelled as 1a and 1b.

| Species | Met-Elongator (1) | Met-Initiator (2) | Ile(CAU) (3) |
|---|---|---|---|
| *Crinipellis perniciosa* | 44802-44874 | 14445-14516 | 62938-63011 |
| *Schizophyllum commune* | 16251-16324 | 18809-18881 | 23325-23398 |
| *Schizosaccharomyces japonicus* | 42182-42255 | 63581-63652 | 48100-48171 |
| *Schizosaccharomyces octosporus* | 27317-27391 | 40910-40983 | |
| *Schizosaccharomyces pombe* | 4591-4665 | 14081-14153 | 17542-17613 |
| *Candida stellata* | 18882-18954 | 22272-22343 | 22191-22262 |
| *Pichia canadensis* | 21635-21709 | 19011-19083 | 19088-19158 |
| *Kluyveromyces lactis* | 6768-6840 | 14545-14616 | |
| *Kluyveromyces thermotolerans* | 21284-21356 | 7352-7424 | |
| *Ashbya gossypii* | 5758-5828 | 18723-18794 | |
| *Candida glabrata* | 13-86 | 6889-6959 | |
| *Saccharomyces cerevisiae* | 72630-72705 | 85035-85112 | |
| *Saccharomyces castellii* | 3292-3364 | 8530-8603 | |
| *Saccharomyces servazzii* | 10527-10600 | 15211-15284 | |
| *Hypocrea jecorina* | 16538-16609 | 14475-14545 | 14691-14763 |
| *Lecanicillium muscarium* | 5774-5845 | 4898-4968 | 5116-5188 |
| *Penicillium marneffei* | 23057-23127 | 22140-22210 | 22214-22286 |
| *Podospora anserina* | (1a) 11266-11336<br>(1b) 26071-26141 | 9600-9670 | 9437-9509 |
| *Allomyces macrogynus* | 42129-42203 | 47535-47605 | 44168-44241 |
| *Candida albicans* | | 33486-33557 | 26935-27005 |
| *Candida parapsilosis* | | 21711-21780 | 21792-21862 |
| *Yarrowia lipolytica* | | 3272-3343 | 3374-3445 |